\documentclass[12pt,a4paper]{article}
\pdfoutput=1

\usepackage{amsmath,amsfonts,color,latexsym}
\usepackage{amssymb,a4wide,epsfig}

\usepackage{cite}
\usepackage{array}

\def \be  {\begin{equation}}
\def \ee  {\end{equation}}
\def \ba  {\begin{eqnarray}}
\def \ea  {\end{eqnarray}}
\def \bb  {\begin{thebibliography}}
\def \eb  {\end{thebibliography}}
\def \lab #1 {\label{#1}}

\newcommand\PN{\mathbb{PN}}
\newcommand\cA{\mathcal{A}}

\newcommand\cM{\mathcal{M}}
\newcommand\cN{\mathcal{N}}

\newcommand\cO{\mathcal{O}}

\newcommand\cS{\mathcal{S}}

\newcommand\cW{\mathcal{W}}
\newcommand\cY{\mathcal{Y}}
\newcommand\cZ{\mathcal{Z}}
\newcommand\cR{\mathcal{R}}

\newcommand\MHV{\mathrm{MHV}}

\newcommand\C {\mathbb{C }}
\newcommand\R {\mathbb{R }}

\newcommand\CP {\mathbb{CP}}

\newcommand\rd{\mathrm{d}}
\newcommand\rD{\mathrm{D}}

\newcommand\im{\mathrm{i}}
\newcommand\rU{\mathrm{U}}
\newcommand\rW{\mathrm{W}}
\newcommand\la{\langle}
\newcommand\ra{\rangle}
\newcommand\del{\partial}
\newcommand\delbar{\bar{\partial}}

\newcommand{\bea}{\begin{eqnarray}\label}
\newcommand{\eea}{\end{eqnarray}}

\newcommand{\captionfonts}{\small}

\makeatletter  % Allow the use of @ in command names
\long\def\@makecaption#1#2{%
  \vskip\abovecaptionskip
  \sbox\@tempboxa{{\captionfonts #1: #2}}%
  \ifdim \wd\@tempboxa >\hsize
    {\captionfonts #1: #2\par}
  \else
    \hbox to\hsize{\hfil\box\@tempboxa\hfil}%
  \fi
  \vskip\belowcaptionskip}
\makeatother   % Cancel the effect of \makeatletter

\begin{document}
  
\thispagestyle{empty}

\vskip1.5truecm
\begin{center}
\vskip 0.2truecm

 {\Large\bf
Descent equations for superamplitudes
}
\vskip 1truecm
%\vfill

%\vskip 1truecm
%\vfill
{\bf    Mathew Bullimore$^{\sharp}$ and  David Skinner$^{\natural}$\\
}

\vskip 0.5truecm

$^{\sharp}${\it  Rudolf Peierls Centre for Theoretical Physics,\\
	1 Keble Road, Oxford, OX1 3NP, UK\\}
 \vskip .2truecm
$^{\natural}${\it Perimeter Institute for Theoretical Physics,\\ 
	31~Caroline~St., Waterloo, ON, N2L 2Y5, 
	Canada\\}
\vskip 1truecm

\end{center}

\centerline{\bf Abstract}
\medskip

At loop level in planar $\cN=4$ SYM, the dual superconformal symmetry of tree amplitudes is lost. This is true even if one uses a 
supersymmetry preserving regulator, and even for finite quantities that remain dual conformally invariant. We examine this 
breaking from the dual point of view of the super Wilson Loop, tracing it to the difference between supersymmetries of the self-dual and 
of the full theories. We show that the anomaly is controlled by a descent equation that determines the derivative of an $\ell$-loop 
amplitude in terms of a single non-trivial integral of an $(\ell\!-\!1)$-loop amplitude. We propose that this equation can be used 
recursively to construct  multi-loop amplitudes in a way that makes their transcendentality manifest.

\newpage

%%%%%%%%%%%%%%%%%%%%%%%%%%%%%%%%%%%%%%%%%%%%%%%%%%%
%%%%%%%%%%%%%%%%%%%%%%%%%%%%%%%%%%%%%%%%%%%%%%%%%%%

\section{Introduction}
\label{sec:intro}

Tree level scattering amplitudes in $\cN=4$ SYM are invariant under the action of the infinite dimensional Yangian algebra 
$\cY[\frak{psu}(2,2|4)]$~\cite{Drummond:2008vq,Drummond:2009fd}. This statement is highly constraining: the tree amplitudes are 
completely determined by this Yangian, together with knowledge of their behaviour in collinear limits~\cite{Korchemsky:2009hm,Bargheer:2009qu}.

At loop level, much of this symmetry is broken. Broken symmetries can still place powerful constraints on the 
amplitudes, provided the structure of the breaking is understood. For example, the infra-red divergences of scattering amplitudes that 
violate conformal invariance take a universal exponential form~\cite{Magnea:1990zb,Sterman:2002qn}, and demanding consistency 
with this structure played a key role in the construction of the Bern-Dixon-Smirnov ansatz~\cite{Bern:2005iz} for all-loop planar MHV 
amplitudes. This ansatz also provides a particular solution to the anomalous Ward identity for {\it dual} special conformal 
transformations~\cite{Drummond:2007au} that follows from the duality between planar scattering amplitudes and null polygonal Wilson 
Loops. The Ward identity then states that, once the BDS ansatz is factored out, any remainder must be a dual  conformal invariant.

\medskip

In this paper, we will be concerned with the dual \emph{supersymmetry} of scattering amplitudes, discovered at tree level 
in~\cite{Drummond:2008vq}. All known $\ell$ loop N$^k$MHV  amplitudes in planar $\cN=4$ SYM may 
be written as
\be
	\cM  = \sum \left(\hbox{leading singularity}\right) \, \times \,\left(\hbox{bosonic integral}\right)
\label{loopschematic}
\ee
where the leading singularities are Yangian invariants~\cite{Bullimore:2009cb,Kaplan:2009mh,Drummond:2010uq,Arkani-Hamed:2010kv} and the bosonic integrals generally require regularization. If the external data for the scattering process is given in terms of momentum supertwistors~\cite{Hodges:2009hk,Mason:2009qx}, then the superconformal supercharges $Q$ and $\bar S$ may be represented by the first order differential operators
\be
	Q_{\rm ext}= \sum_{i=1}^n\lambda_{i}\frac{\del}{\del\chi_i}
	\qquad\hbox{and}\qquad
	\bar S_{\rm ext}=  \sum_{i=1}^n\mu_i\frac{\del}{\del\chi_i}\ .
\label{succeeds}
\ee
acting on this unconstrained data. It is immediately clear that these generators annihilate any $\cM$ of the form~\eqref{loopschematic}, 
since they annihilate both the leading singularities and the purely bosonic integrals.  However, the conjugate supercharges
\be
	\bar Q_{\rm ext} = \sum_{i=1}^n\chi_i\frac{\del}{\del\mu_i}
	\qquad\hbox{and}\qquad
	S_{\rm ext} = \sum_{i=1}^n\chi_i\frac{\del}{\del\lambda_{i}}
\label{fails}
\ee
fail to annihilate the loop integrals since they differentiate with respect to the bosonic variables. Thus, as was observed in~\cite{Korchemsky:2009hm}, these $\bar Q$ and $S$ supersymmetries are broken at the quantum level.

\medskip

There are a number of reasons why this $\bar Q$ anomaly seems particularly puzzling. Firstly, the breaking of 
$\bar Q$  apparently has nothing to do with the fact that loop amplitudes require regularization. Even finite quantities, such as the 
remainder function or the ratio $\cM/\cM_{\rm MHV}$ of the superamplitude to the MHV amplitude, fail to be annihilated by~\eqref{fails},
despite being fully dual conformally invariant. Furthermore, from the point of view of the duality to Wilson Loops, these dual 
superconformal charges are just the ordinary superconformal charges of the dual theory.  As stressed in~\cite{Korchemsky:2009hm}, we 
usually expect to be able to find a regularization scheme that preserves Poincar{\'e} supersymmetry $Q$ and $\bar Q$, rather than the 
chiral half of the superalgebra consisting of $Q$ and $\bar S$ that is preserved here.

Secondly, instead of stripping off the MHV tree amplitude and passing to momentum twistor space, one could equally choose to remove 
a factor of the $\overline{\MHV}$ tree amplitude and work in dual momentum twistor space ({\it i.e.} the dual projective space). Then the 
r{\^o}les would be reversed: for the same reasons as above, one would find that $Q$ and $\bar S$ (represented by $\bar\chi\del/\del\bar Z$) fail to annihilate the resulting expression, while $\bar Q$ and $S$  ($\sim \bar Z\del/\del\bar \chi$) would be preserved.

These points strongly suggest that the failure of~\eqref{fails} to annihilate loop amplitudes is strongly tied to the representation of 
scattering amplitudes in a (dual) chiral superspace. To try to circumvent this, in~\cite{Caron-Huot:2011ky}, Caron-Huot constructed a 
non-chiral extension of the super Wilson Loop that is (dual) supersymmetric\footnote{At least for Poincar{\'e} 
supersymmetry. The superconformal algebra constrains $\{Q,\bar S\} = K$, so any $Q$-invariant process with a $K$ anomaly must also 
be anomalous under $\bar S$.} (see also~\cite{Ooguri:2000ps}). However, while this non-chiral super Wilson Loop is 
undoubtedly a fascinating object in its own right, it is no longer dual purely to the amplitudes:  its $\bar\theta_i$ expansion involves a 
large number of additional terms that are responsible for restoring $\bar Q$ symmetry, but whose independent meaning is not clear.

\medskip

In this paper, we show that the $\bar Q$ anomaly can be understood purely within the context of the chiral superloop / superamplitude 
duality. More precisely, working in the Wilson Loop context and treating $\bar Q$ as a regular supersymmetry, in section~\ref{sec:susytrans} we show that the generators~\eqref{fails} correspond to field transformations that are symmetries of the 
\emph{self-dual} sector of $\cN=4$ SYM only; they do not even preserve the classical action of full $\cN=4$ SYM. In 
section~\ref{sec:Ward} we propose a Ward identity that states that the full (and completely standard) $\bar Q$ transformations are 
indeed symmetries of the all-orders chiral superloop; the failure of~\eqref{fails} to annihilate the superloop is compensated by the action 
of the difference 
\be
	\bar Q^{(1)}\equiv \bar Q_{\rm full}-\bar Q^{(0)}
\label{Qbar1def}
\ee
between the supercharges in the full and self-dual theories. In section~\ref{sec:1loopMHV} we perform a simple test of this Ward identity, 
using it to reconstruct (the symbol of) the 1-loop MHV amplitude.

We believe this interpretation of the $\bar Q$ anomaly is very natural from the point of view of the super Wilson Loop. From the point of 
view of the scattering amplitude however, we will see that the Ward identity mixes different orders in perturbation theory. Thus, although 
the complete planar S-matrix is fully dual supersymmetric, this is not true of individual $\ell$-loop N$^k$MHV amplitudes. In more detail, 
the $\bar Q$ non-invariance of an $\ell$-loop N$^k$MHV amplitude will be seen to be corrected by a term coming from the 
$(\ell-1)$-loop N$^{k+1}$MHV amplitude. (Note that understanding the $\bar Q$ anomaly thus requires the superloop / superamplitude 
duality; it cannot be seen purely within the duality between bosonic Wilson Loops and MHV amplitudes).

In section~\ref{sec:descent} we investigate the structure of the Ward identity in more detail. We show that 
we can view $\bar Q_{\rm ext}$ and $\bar Q^{(1)}$ as generating a descent procedure that governs the structure of the 
$\bar Q_{\rm ext}$ anomaly. In this procedure, $k$ and $\ell$ play the roles of ghost number and form degree. The $\bar Q^{(1)}$ action 
can indeed be understood as an operation carried out on a superloop with one extra vertex, taken in a particular collinear limit. The 
resulting descent equation powerfully constrains the form of multiloop N$^k$MHV amplitudes: the $\bar Q_{\rm ext}$ variation -- and 
hence the first order derivative -- of higher loop amplitudes is determined in terms of a single integral of lower loop ones. 

It is quite remarkable that only a {\it single} (non-trivial) integral is involved. To increase the loop order by one we usually expect to have 
to perform a four-dimensional integral over another loop momentum, or Wilson Loop vertex. However, there is much redundancy in this 
description and in fact all known $\ell$-loop amplitudes in planar $\cN=4$ SYM obey an extension of the Kotikov - Lipatov 
principle~\cite{Kotikov:2002ab} for the cusp anomalous dimension, which states that they have transcendentality $2\ell$; {\it i.e.}, 
they can be expressed as only $2\ell$-fold iterated integrals of rational functions. The descent equations presented here make this 
transcendentality manifest.

The fact that the complete S-matrix is invariant under dual supersymmetry while individual amplitudes are not is strongly reminiscent of 
work of Korchemsky \& Sokatchev~\cite{Korchemsky:2009hm}, of Sever \& Vieira~\cite{Sever:2009aa} and of Beisert 
{\it et al.}~\cite{Bargheer:2009qu,Beisert:2010gn}. These authors give a careful study of the action of the original superconformal 
generators on the scattering amplitude, and show that certain generators should be `corrected' to ensure the tree amplitude remains 
invariant even when external states become collinear. Essentially the same corrections account for the violation of superconformal 
symmetry at loop level, once collinear singularities between external and internal states are considered. Since the dual Poincar{\'e} 
supercharge $\bar Q$ considered here coincides~\cite{Drummond:2008vq} with the original superconformal supercharge 
$\bar s\sim \eta\del/\del\bar\lambda$ that receives corrections in the collinear limit, one suspects there must be a close connection 
between the story here and that of~\cite{Korchemsky:2009hm,Sever:2009aa,Bargheer:2009qu,Beisert:2010gn}. We finish by elucidating 
this relation in section~\ref{sec:exacting}.

\medskip

\emph{Note added}: While this paper was in preparation, we became aware of the work~\cite{SimonandSong} by Simon Caron-Huot and Song He, which has some overlap with the work presented here.

%%%%%%%%%%%%%%%%%%%%%%%%%%%%%%%%%%%%%%%%%%%%%%%%%%%
%%%%%%%%%%%%%%%%%%%%%%%%%%%%%%%%%%%%%%%%%%%%%%%%%%%

\section{Supersymmetries of self-dual $\cN=4$ SYM}
\label{sec:susytrans}

In~\cite{Mason:2010yk,Caron-Huot:2010ek,Bullimore:2011ni} the duality between scattering amplitudes (divided by the MHV tree) and 
null polygonal Wilson Loops in planar $\cN=4$ SYM was extended beyond the MHV sector to the full superamplitude at the level of the 
four-dimensional integrand. This was achieved by considering the correlation function\footnote{We use conventions in which the 
(bosonic) covariant derivative ${\rm D}={\rm d}-\im A$.}
\be
	\rW[C] \equiv \frac{1}{N} \left\la {\rm Tr\,P}\exp\left(\im\oint_C {\mathbb A}\right)\right\ra
\label{superloop}
\ee
of the trace of the holonomy in the fundamental representation of a certain connection\footnote{Here, $a=1,\ldots,4$ indexes the R-symmetry, while $\alpha=0,1$ and $\dot\alpha=\dot0,\dot1$ are left and right Weyl spinor indices.}
\be
	{\mathbb A}(x,\theta) = A_{\alpha\dot\alpha}(x,\theta)\rd x^{\alpha\dot\alpha} + \Gamma_{\alpha a}(x,\theta)\rd \theta^{\alpha a}
\label{superconnection}
\ee	
in chiral superspace, whose detailed form will be given later. The super Wilson Loop in~\eqref{superloop} is computed around a curve $C$ that is the lift 
\be
	x_i(t) = x_i + t(x_{i+1}-x_i)
		\qquad\qquad 
	\theta_i(t) = \theta_i + t(\theta_{i+1}-\theta_i)
\label{contour}
\ee
of the null polygonal contour to chiral superspace, where 
\be
	x_{i+1}-x_i=\lambda_i\bar\lambda_i\qquad \hbox{and}\qquad
	\theta_{i+1}-\theta_i=\lambda_i\eta_i\ .
\label{nullconstraints}
\ee
Expanding~\eqref{superloop} in powers of the fermionic coordinates $\eta_i$ allows for arbitrary external helicities in the 
scattering process. The calculation is in fact most easily carried out using a twistor formulation of 
both the operator and the $\cN=4$ action~\cite{Boels:2006ir}, because the twistor contour solves the constraints~\eqref{nullconstraints} 
automatically. In particular, it has been shown that BCFW recursion relations for the scattering amplitude~\cite{Britto:2005fq,Arkani-Hamed:2010kv} follow from the loop equations for this super Wilson Loop~\cite{Bullimore:2011ni}.

\medskip

One of the most striking aspects of this duality is that the complete \emph{tree-level} superamplitude was found to arise from taking the 
Wilson Loop correlator $\la {\rm W}\ra_{\rm sd}$ in \emph{self-dual} $\cN=4$ SYM only. In space-time, this theory is given by the 
action~\cite{Siegel:1992za,Chalmers:1996rq}
\be
	S_{\rm sd}
	=\frac{1}{{\rm g}^2}\int\rd^4x\  {\rm Tr}\left( G^{\alpha\beta}F_{\alpha\beta}
	+2\im\bar\psi_{\dot\alpha a} D^{\alpha\dot\alpha}\psi_\alpha^{\ a} 
	+\frac{1}{2} D_\mu\phi^{ab}D^\mu\phi_{ab} 
	+ \bar\psi^{\dot\alpha}_{\ a}\left[\phi^{ab}, \bar\psi_{\dot\alpha b}\right]\right)
\label{sdYMaction}
\ee
where $G_{\alpha\beta} = G_{\beta\alpha}$ represents an anti self-dual two-form $G$ in the adjoint representation of the gauge group. 
$G$ is an auxiliary field whose equation of motion $F_{\alpha\beta}=0$ constrains the anti self-dual part of the Yang-Mills fieldstrength 
$F={\rm d}A-\im[A,A]$ to vanish. Self-dual Yang-Mills is an integrable theory in four dimensions~\cite{Ward:1977ta} and so has no 
scattering amplitudes. It is therefore quite remarkable that it nonetheless knows about the complete classical S-matrix of full Yang-Mills 
via the correlation function~\eqref{superloop}.

The self-dual action~\eqref{sdYMaction} possesses $\cN=4$ superconformal symmetry. In particular, focussing on the Poincar{\'e} 
supersymmetries $\delta_{\rm sd} = \epsilon Q + \bar\epsilon\bar Q$, it is straightforward to check that~\eqref{sdYMaction} is invariant 
under the field transformations~\cite{Siegel:1992za,Siegel:1992wd,Belitsky:2011zm}
\be
\begin{aligned}
	\delta_{\rm sd} A &= -\im|\epsilon^a\ra[\bar\psi_a|
	\qquad&\phantom{+}&\\
	\delta_{\rm sd}|\bar\psi_a] &= \ D\phi_{ab} |\epsilon^b\ra
	&+&\quad
	\im[\bar\epsilon_a| F^+ \\
	\delta_{\rm sd}\phi_{ab} &=-\im\varepsilon_{abcd}\la\epsilon^c\psi^d\ra
	&+&\quad
	\im([\bar\epsilon_a\bar\psi_b] - [\bar\epsilon_b\bar\psi_a])\\
	\delta_{\rm sd}|\psi^a\ra &= \im G|\epsilon^a\ra+\frac{\im}{2}\left[\phi^{ab},\phi_{bc}\right]|\epsilon^c\ra
	&+&\quad
	[\bar\epsilon_b| D\phi^{ab}\\
	\delta_{\rm sd} G_{\alpha\beta} &= \epsilon^a_{\ (\alpha}\left[\psi^b_{\beta)},\phi_{ab}\right]
	&+&\quad
	[\bar\epsilon_a|D_{(\alpha}\psi_{\beta)}^{\ a}\quad .
\end{aligned}
\label{susy1}
\ee
These transformations leave the action invariant (up to total derivatives), but as usual represent the supersymmetry algebra only up to the field equations of~\eqref{sdYMaction} and field dependent gauge transformations. 

\medskip

From the purposes of this paper, the most important feature of the transformations~\eqref{susy1}  is that they 
do \emph{not} coincide with the field transformations that generate supersymmetries of full (non self-dual) $\cN=4$ SYM. 

\medskip

More precisely, if the action of full $\cN=$ SYM is written in Chalmers \& Siegel form
\be
	S_{\rm full} = S_{\rm sd} + S_{\rm MHV}
\label{action}
\ee
with $S_{\rm sd}$ as in~\eqref{sdYMaction} and
\be
	S_{\rm MHV} =\frac{1}{{\rm g}^2}\int\rd^4x\  {\rm Tr}\left( -\frac{1}{2}G^{\alpha\beta}G_{\alpha\beta} 
	 +\psi^{\alpha a}[\phi_{ab}, \psi_\alpha^{\ b}] + \frac{1}{8} [\phi^{ab}, \phi^{cd}][\phi_{ab},\phi_{cd}] \right)\ ,\label{MHVaction}
\ee
then the $Q$ supersymmetries of the full theory are exactly the same as in~\eqref{susy1} (terms proportional to $\epsilon$). However, for the $\bar Q$ supersymmetries we have
\be
	\bar Q_{\rm full} = \bar Q^{(0)} + \bar Q^{(1)}
\ee
where $\bar Q^{(0)}$ are the transformations of the self-dual theory given in~\eqref{susy1} (terms proportional to $\bar\epsilon$). The 
difference $\bar Q^{(1)}$ acts trivially on $\phi$, $\psi$ and the auxiliary field\footnote{This field remains auxiliary in the full 
theory~\eqref{action}, and is fixed to be the anti self-dual part of the fieldstrength $G_{\alpha\beta}=F_{\alpha\beta}$. The $Q$ 
transformation $\delta G = \epsilon[\psi,\phi]$ agrees with the standard transformation of $F_{\alpha\beta}$ upon using 
the $\psi$ equation of motion, but note that the $Q$ transformations in~\eqref{susy1} remain symmetries of $S_{\rm sd}+S_{\rm MHV}$ without the use of field equations.} $G$, but non-trivially on the gluon and the positive helicity states of the gluino:
\be
\begin{aligned}
	\delta^{(1)} \!A &= \im |\psi^a\ra[\bar\epsilon_a| \\
	\delta^{(1)} |\bar\psi_a] &= -\frac{\im}{2}\,|\bar\epsilon_c]\left[\phi^{cb},\phi^{ab}\right]\ .
\end{aligned}
\label{delta1}
\ee
Since $S_{\rm MHV}=S_{\rm MHV}[\phi, \psi, G]$, we immediately see that $\delta^{(1)}S_{\rm MHV}=0$. Invariance of the self-dual action under the self dual supersymmetries ($\delta_{\rm sd}S_{\rm sd}=0$) and of the full action under the full supersymmetries ($\delta_{\rm full}S_{\rm full}=0$) then implies that
\be
	\delta_{\rm sd} S_{\rm MHV} + \delta^{(1)}S_{\rm sd} = 0
\ee
One may verify that only this combination - rather than the individual terms - vanishes. In other words, self-dual $\cN=$ Yang-Mills is not
invariant under the supersymmetries of the full theory, nor is the full theory invariant under the supersymmetries of the self-dual theory.

%%%%%%%%%%%%%%%%%%%%%%%%%%%%%%%%%%%%%%%%%%%%%%%%%%%
%%%%%%%%%%%%%%%%%%%%%%%%%%%%%%%%%%%%%%%%%%%%%%%%%%%

\section{Ward Identities for super Wilson Loops}
\label{sec:Ward}

In the rest of the paper, we will show that the difference between the self-dual and full $\bar Q $ supersymmetries is responsible 
for the anomaly in this symmetry for loop amplitudes. The reason the difference between self-dual and full supersymmetries is 
related to the difference between (dual) supersymmetries of tree and loop amplitudes is that, while the self-dual correlator 
$\rW_{\rm sd}$ corresponds to the tree-level S-matrix, to obtain quantum corrections to the scattering amplitude, we 
must  instead compute the super Wilson Loop correlator in the full theory. In particular, every time one calls upon 
$S_{\rm MHV}$ to provide a vertex for diagrams contributing to this full correlator, the loop order of the corresponding amplitude 
calculation is increased -- these are chiral Lagrangian insertions in the language of~\cite{Eden:2010ce}. Calling upon $S_{\rm MHV}$ a total of $\ell$ times yields a contribution to the superloop corresponding to a piece of the $\ell$-loop scattering amplitude. In twistor space, $S_{\rm MHV}$ becomes an infinite sum of MHV vertices~\cite{Boels:2006ir}. The axial gauge Feynman diagrams of the twistor Wilson Loop are 
the planar duals of MHV diagrams for the scattering amplitude~\cite{Mason:2010yk} while including the effect of these vertices in the 
loop equations~\cite{Bullimore:2011ni} generates the correction to the tree-level BCFW recursion relations, promoting them to the 
all-loop integrand recursion relation of~\cite{Arkani-Hamed:2010kv}.

\medskip

Now, when one acts on the external data of a Wilson Loop correlator with the operator 
\be
	\bar{Q}_{\rm ext} 
	= \sum_i \theta_i\frac{\del}{\del x_i} + \eta_i\frac{\del}{\del\bar\lambda_i}
	= \sum_i \chi_i\frac{\del}{\del \mu_i}
\label{Qbarext}
\ee
(either on chiral superspace-time or in twistor space), it is important to understand which $\bar Q$ this corresponds to. The 
choice is easy: since $\bar Q_{\rm ext}$ annihilates tree amplitudes and since these are computed by the expectation value of the super 
Wilson Loop in the \emph{self-dual} theory only, $\bar Q_{\rm ext}$ must act on the fields as the self-dual transformations. For only then 
do we have the Ward identity
\be
	\sum_i\chi_i\frac{\del }{\del\mu_i} \rW[C_n]
	= \frac{1}{N} \left\la\left[\bar Q^{(0)}\ ,  {\rm Tr\,P}\exp\left(\im\oint_{C_n} {\mathbb A}\right) \right]\right\ra_{\rm sd} = 0 
\label{WIselfdual}
\ee
since it is $\bar Q^{(0)}$, and not the full $\bar Q$, that generates a symmetry of the self-dual theory.

To check that~\eqref{Qbarext} really does generate only the self-dual supersymmetries, note that the geometric action $\cZ^I\del/\del \cZ^J$ of the superconformal group on supertwistor space generates an action on the twistor superfield
\be
	\mathcal{A}(Z,\chi)= a(Z) + \chi^a\,\tilde\Psi_a(Z) + \frac{1}{2!}\chi^a\chi^b\,\Phi_{ab}(Z)
	+\frac{\varepsilon_{abcd}}{3!}\chi^a\chi^b\chi^c\,\Psi^d(Z)+\frac{\varepsilon_{abcd}}{4!}\chi^a\chi^b\chi^c\chi^d\,g(Z)
\label{twistorsuperfield}
\ee
in the usual way
\be
	\delta\cA =  \epsilon^I_{\ J}\cZ^I\frac{\del\cA}{\del\cZ^J} \ .
\label{Liederivative}
\ee
by Lie derivation along $V=\epsilon^I_{\ J}\cZ^I\del/\del \cZ^J$. These are manifest symmetries of the holomorphic Chern-Simons action~\cite{Witten:2003nn}
\be
	S = \frac{1}{{\rm g}^2}\int_{\mathbb{CP}^{3|4}}\hspace{-0.2cm}{\rm D}^{3|4}Z\wedge
	{\rm Tr}\left(\cA\wedge\delbar\cA + \frac{2\im}{3}\cA\wedge\cA\wedge\cA\right)
\label{hCS}
\ee
that corresponds to the self-dual action~\eqref{sdYMaction} on twistor space. In particular, the Poincar{\'e} $\bar Q$ transformations act 
on the twistor fields as 
\be
\begin{aligned}
	\delta a(Z) &= 0\qquad\qquad \qquad
	&\delta \tilde\Psi_a(Z) &= \bar\epsilon^{\dot\alpha}_{\ a} \frac{\del a(Z)}{\del\mu^{\dot\alpha}} &\\
	\delta\Phi_{ab}(Z)&=\bar\epsilon^{\dot\alpha}_{\ [a}\frac{\del\tilde\Psi_{b]}(Z)}{\del\mu^{\dot\alpha}}
	&\delta\Psi^a(Z) &=  \bar\epsilon^{\dot\alpha}_{\ b}\frac{\del\Phi^{ab}(Z)}{\del\mu^{\dot\alpha}}\qquad\qquad
	&\delta g(Z) = \bar\epsilon^{\dot\alpha}_{\ a}\frac{\del\Psi^a(Z)}{\del\mu^{\dot\alpha}}
\end{aligned}
\label{twistorsusy}
\ee
so that the lowest component $a(Z)$ is left invariant. Under the (Abelian\footnote{The transformations~\eqref{twistorsusy} are the twistor space transformations of the component fields even in the non-Abelian case. The non-linearities in the space-time transformations~\eqref{susy1} arise from non-linearities in the non-Abelian Penrose transform.}) Penrose transform, the self-dual part of the space-time 
fieldstrength is
\be
	F_{\dot\alpha\dot\beta}(x) = \oint \la\lambda\rd\lambda\ra \left.\frac{\del^2 a}{\del\mu^{\dot\alpha}\del\mu^{\dot\beta}}\right|_{\mu=x\lambda}
\ee
and so
\be
	\delta F_{\dot\alpha\dot\beta}(x) = \oint \la\lambda\rd\lambda\ra
	\left.\frac{\del^2 \delta a}{\del\mu^{\dot\alpha}\del\mu^{\dot\beta}}\right|_{\mu=x\lambda}
	=0\ .
\ee
This is in agreement with~\eqref{susy1}, but is incompatible with~\eqref{delta1}. Therefore, the geometric transformation~\eqref{Qbarext} 
of the external twistor data indeed generates the field transformations that are supersymmetries of only the self-dual theory.

\medskip

Now, if we act with the same operator~\eqref{Qbarext} on the external data of a Wilson Loop correlator in full $\cN=4$ SYM -- {\it i.e.}, including loop corrections to the amplitude -- then the Ward identity~\eqref{WIselfdual} receives a correction, becoming
\be
\begin{aligned}
	\sum_i\chi^i\frac{\del }{\del\mu_i} \rW [C_n]
	&= \frac{1}{N} \left\la\left[\bar Q^{(0)},  {\rm Tr\,P}\exp\left(\im\oint_{C_n} {\mathbb A}\right) \right]\right\ra_{\rm full} \\
	&=  -\frac{1}{N} \left\la\left[\bar Q^{(1)},  {\rm Tr\,P}\exp\left(\im\oint_{C_n} {\mathbb A}\right) \right]\right\ra_{\rm full} 
\label{Wardid}
\end{aligned}
\ee
reflecting the fact that it is $\bar Q_{\rm full}= \bar Q^{(0)}+\bar Q^{(1)}$ that generates a symmetry of the full theory. The non-zero right 
hand side of~\eqref{Wardid} measures the failure of the full $\cN=4$ action to be invariant under chiral supersymmetry transformations. 

\medskip

Equation~\eqref{Wardid} is one of the main claims of this paper. It is simply the assertion that the correlator of the super Wilson Loop in 
the full quantum theory is invariant under $\cN=4$ Poincar{\'e} supersymmetry. Equivalently, all-loop scattering amplitudes are exactly 
invariant under dual Poincar{\'e} supersymmetry. However, beyond tree level these dual supersymmetries are not generated by the straightforward action of~\eqref{Qbarext}.

\medskip

In the following sections, we will examine the structure of~\eqref{Wardid} more closely and test it in a few simple examples. We will see 
that it provides a straightforward route to compute the symbol of loop level scattering amplitudes directly, without recourse to the 
integrand. Let us first conclude this section with a few clarifying remarks. Firstly, the fact that loop corrections to the scattering amplitude 
come only from vertices drawn from $S_{\rm MHV}$ suggests we rescale the fields so that the action becomes 
\be
	S_{\rm full} = S_{\rm sd}+{\rm g}^2S_{\rm MHV}\ ,
\label{rescaledaction}
\ee
with $S_{\rm sd}$ and $S_{\rm MHV}$ now being independent of the coupling constant. The required rescalings are uniquely determined to be
\be
	A\to A\qquad 
	|\bar\psi_a]\to{\rm g}^{\frac{1}{2}} |\bar\psi_a]\qquad 
	\phi_{ab}\to {\rm g}\,\phi_{ab}\qquad 
	\la\psi^a|\to {\rm g}^{\frac{3}{2}}\la\psi^a|\qquad 
	G\to {\rm g}^2G\ .
\label{rescalefields}
\ee
We also rescale $|\theta^a\ra\to {\rm g}^{-\frac{1}{2}}|\theta^a\ra$ to ensure that the superconnection $\mathbb{A}$ itself remains 
independent of the coupling. With this normalization, which was used in~\cite{Mason:2010yk,Caron-Huot:2010ek,Bullimore:2011ni}, the perturbative 
expansion of the super Wilson Loop correlator matches that of the amplitude order by order in ${\rm g}^2$. Having rescaled $\theta$, we also rescale $\bar\theta\to{\rm g}^{\frac{1}{2}}\bar\theta$ so that $x+\theta\bar\theta$ is unchanged. If we finally perform a compensating rescaling
\be 
	\la\epsilon^a|\to {\rm g}^{-\frac{1}{2}}\la\epsilon^a| \qquad\hbox{and}\qquad |\bar\epsilon_a]\to{\rm g}^{\frac{1}{2}}|\bar\epsilon_a]
\label{paramrescale}
\ee
in the parameters of the supersymmetry transformations, we find that the self-dual transformations of equation~\eqref{susy1} remain independent of the coupling constant, while the transformations of~\eqref{delta1} become proportional to ${\rm g}^2$:
\be
\begin{aligned}
	\delta^{(1)} \!A &= \im{\rm g}^2 |\psi^a\ra[\bar\epsilon_a| \\
	\delta^{(1)} |\bar\psi_a] &= -\frac{\im}{2}{\rm g^2}\,|\bar\epsilon_c]\left[\phi^{cb},\phi_{ba}\right]\ .
\end{aligned}
\label{delta1coupling}
\ee
With this normalization, the Ward identity~\eqref{Wardid} becomes 
\be
	\sum_i\left.\chi^i\frac{\del }{\del\mu_i} \rW[C_n] \right|_{\rm g^2}
	=  \left.-\frac{1}{N} \left\la\left[\bar Q^{(1)},  {\rm Tr\,P}\exp\left(\im\oint_{C_n} {\mathbb A}\right) \right]\right\ra\right|_{\rm g^2}
\label{orderbyorder}
\ee
so that dual supersymmetry transformations $\bar Q_{\rm full}$ mix different orders of perturbation theory from the point of view of the 
amplitude. This is the reason the $\bar Q$ anomaly will be useful: the derivative, and hence the symbol\footnote{See {\it e.g.}~\cite{Goncharov:2009pl,Goncharov:2010jf} for 
an introduction to symbols of transcendental functions.}, of higher loop amplitudes may be read off if we understand the $\bar Q^{(1)}$ 
action on lower loop ones.

Secondly, although we have focussed on the anomaly in dual Poincar{\'e} supersymmetry, a similar story is true for the dual 
superconformal symmetry $S^{a\alpha}$. There is again a difference $\delta S^{a\alpha}\equiv S^{a\alpha}_{\rm full}-S^{a\alpha}_{\rm sd}$ between the self-dual and full supercharges, 
and again the self-dual action is invariant only under the self-dual transformations, while the full action is invariant only under the full 
transformations. An important difference between the Poincar{\'e} and conformal supersymmetries is that the Ward 
identity~\eqref{Wardid} does \emph{not} hold for the loop amplitudes, because of collinear / infra-red singularities. Indeed, the 
superconformal algebra enforces
\be
	\{\bar S^{\dot\alpha}_{\ a}\,,\,S^{b\beta}\} = \delta^b_{\ a}K^{\beta\dot\alpha}
	\qquad\hbox{and}\qquad
	[ K^{\beta\dot\beta}\, , \, \bar Q_{\dot\alpha}^{\ a}] = \delta_{\dot\alpha}^{\ \dot\beta} S^{\beta a}\ ,
\label{superconformal}
\ee
so any quantity with a $K$ anomaly -- such as the scattering amplitude -- cannot be invariant even under the action of the full 
superconformal generator. However, since $\bar Q_{\rm full}$ is a symmetry, the second equation in~\eqref{superconformal} shows that the superconformal 
anomaly for $S^{a\alpha}_{\rm full}$ must be governed by a simple supersymmetry transformation of the anomalous Ward identity for 
dual conformal transformations~\cite{Drummond:2007au}. Conversely, quantities that are functions purely of dual conformal cross-ratios (such as the ratio function $\cR\equiv\cM/\cM_{\rm MHV} = \rW_n/(\rW_n|_{\chi=0})$ or the ratios 
\be
	\frac{\la\rW_n\ra\,\la\rW_{\rm sq}\ra}{\la\rW_{\rm top}\ra\la\rW_{\rm bot}\ra}
\ee
of (super) Wilson Loops considered in~\cite{Alday:2010ku,Gaiotto:2011dt,Sever:2011da})  should be fully (dual) superconformal under the action of both $\bar Q_{\rm full}$ and $S_{\rm full}$.

%%%%%%%%%%%%%%%%%%%%%%%%%%%%%%%%%%%%%%%%%%%%%%%%%%%
%%%%%%%%%%%%%%%%%%%%%%%%%%%%%%%%%%%%%%%%%%%%%%%%%%

\section{A simple check in the Abelian case}
\label{sec:1loopMHV}

In this section we perform an explicit 1-loop check of the Ward identity~\eqref{Wardid} for the Abelian theory. We will see 
that computing the rhs of this Ward identity quickly allows one to deduce the 1-loop MHV amplitude. 

In the Abelian case, the only non-trivial $\bar Q^{(1)}$ transformation is
\be
	\delta^{(1)} \!A = \im |\psi^a\ra[\bar\epsilon_a|
\ee
for the photon. Furthermore, the only appearance of $A$ in the Abelian superconnection is in the same place as for a bosonic connection:
\be
	\left.\mathbb{A}(x,\theta)\right|_{\rm Abelian} = A_{\alpha\dot\alpha}\rd x^{\alpha\dot\alpha} + \hbox{terms independent of $A$} \ .
\ee
Using the facts that the adjoint representation is trivial and that holonomy based at some point $x$ is the same as the Wilson Loop operator, the Abelian superloop varies under $\bar Q^{(1)}$ as
\be
	\left[\bar\epsilon\!\cdot\!\bar Q^{(1)}\, , \rW[C_n] \right] 
	= -\frac{1}{N}\left\la\oint [\bar\epsilon_a|{\rm d}x|\psi^a\ra \exp \left(\im\oint \mathbb{A}\right)\right\ra\ ;
\label{insertion}\ee
an insertion of $\Psi$ at some point $x$ on the loop.

\begin{figure}[t]
	\centering
	\includegraphics[height=40mm]{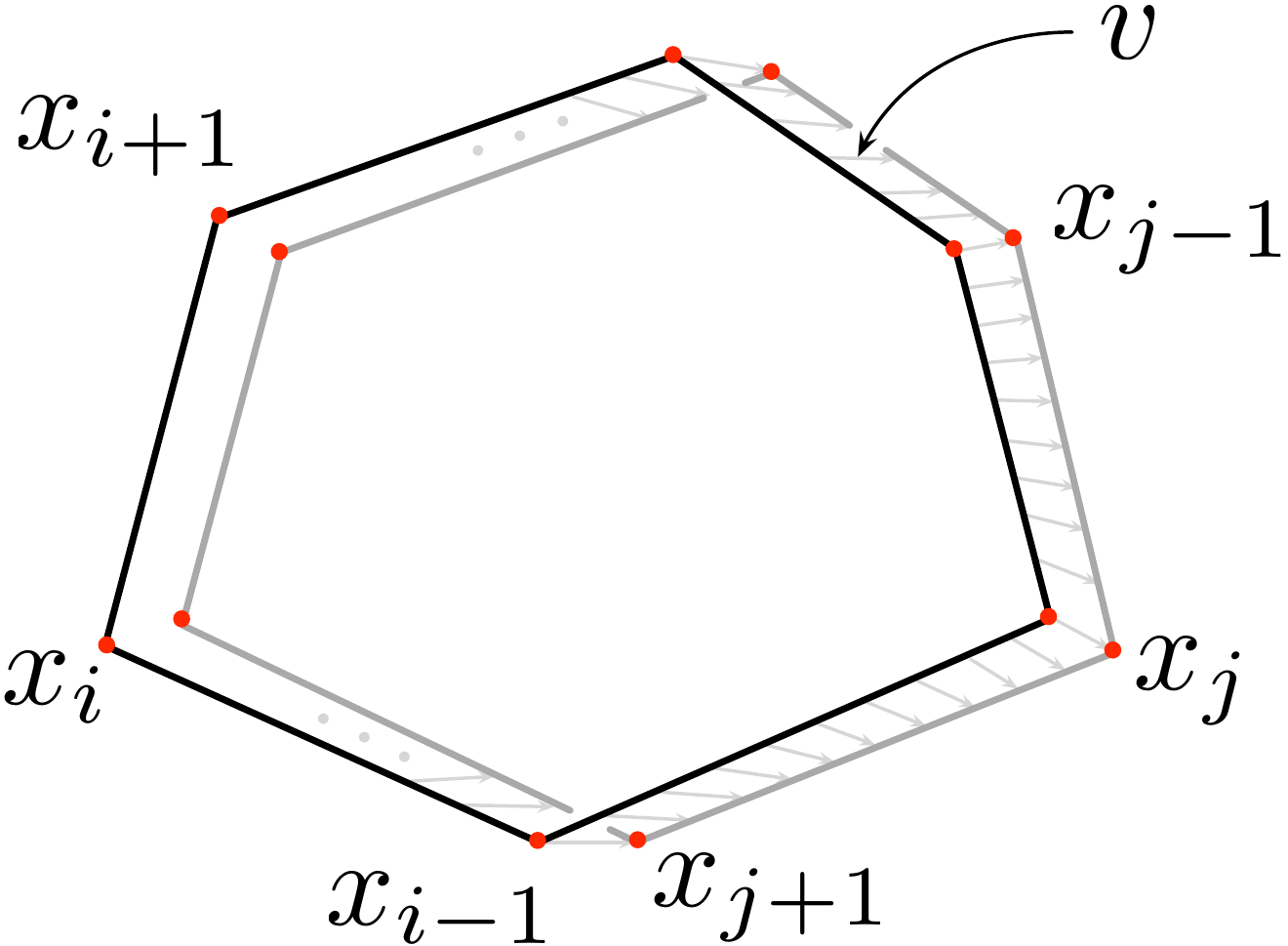}
	\caption{\it\small The framed Wilson Loop. To 1-loop order, we only need consider two copies of the Wilson Loop, obtained by
	translating the original polygon along a nowhere null normal vector field $v$. In our conventions, the vertices of the original Wilson
	Loop are labelled by $\{x_i\}$, whereas the vertices of the translated loop are labelled by $\{x_j\}$. Since $v$ is nowhere null,
	$x_{i,j}^2\neq0$.}
	\label{fig:framing}
\end{figure}

Like the MHV amplitude itself, this correlation function diverges and requires regularization. A convenient way to regularize is by 
framing the loop;  that is, we choose a non-null vector field $v$ normal to $C$ and point split divergent contributions by 
translation along this vector field (see figure~\ref{fig:framing}). An important property of this regularization is that it preserves the 
$Q$ and $\bar S$ supersymmetries of the chiral superloop~\cite{Bullimore:2011ni}. It is closely related to the finite ratios of null 
polygonal Wilson Loops considered by~\cite{Alday:2010ku,Gaiotto:2011dt,Sever:2011da}, since to lowest order in g$^2$, it amounts to 
computing the cross-correlator\footnote{Note that in the non-Abelian theory (relevant for $\ell$-loop N$^k$MHV amplitudes with 
$k+\ell \geq2$) framing regularization does not simply reduce to the cross-correlator~\eqref{framed}.}
\be
	\frac{\left\la\, \rW[C]\,\rW[C']\,\right\ra}{\left\la\rW[C]\right\ra\left\la\rW[C']\right\ra}
\label{framed}
\ee
where $C$ is the original null polygon and $C'$ is the polygon obtained by translating $C$ infinitesimally along $v$.  Because $v$ is non-null, no vertex of $C$ is null-separated from any vertex of $C'$. We make the 
convention that $x_i$ label the vertices of $C$, while $x_j$ label the vertices of $C'$.

\begin{figure}[t]
	\centering
	\includegraphics[height=40mm]{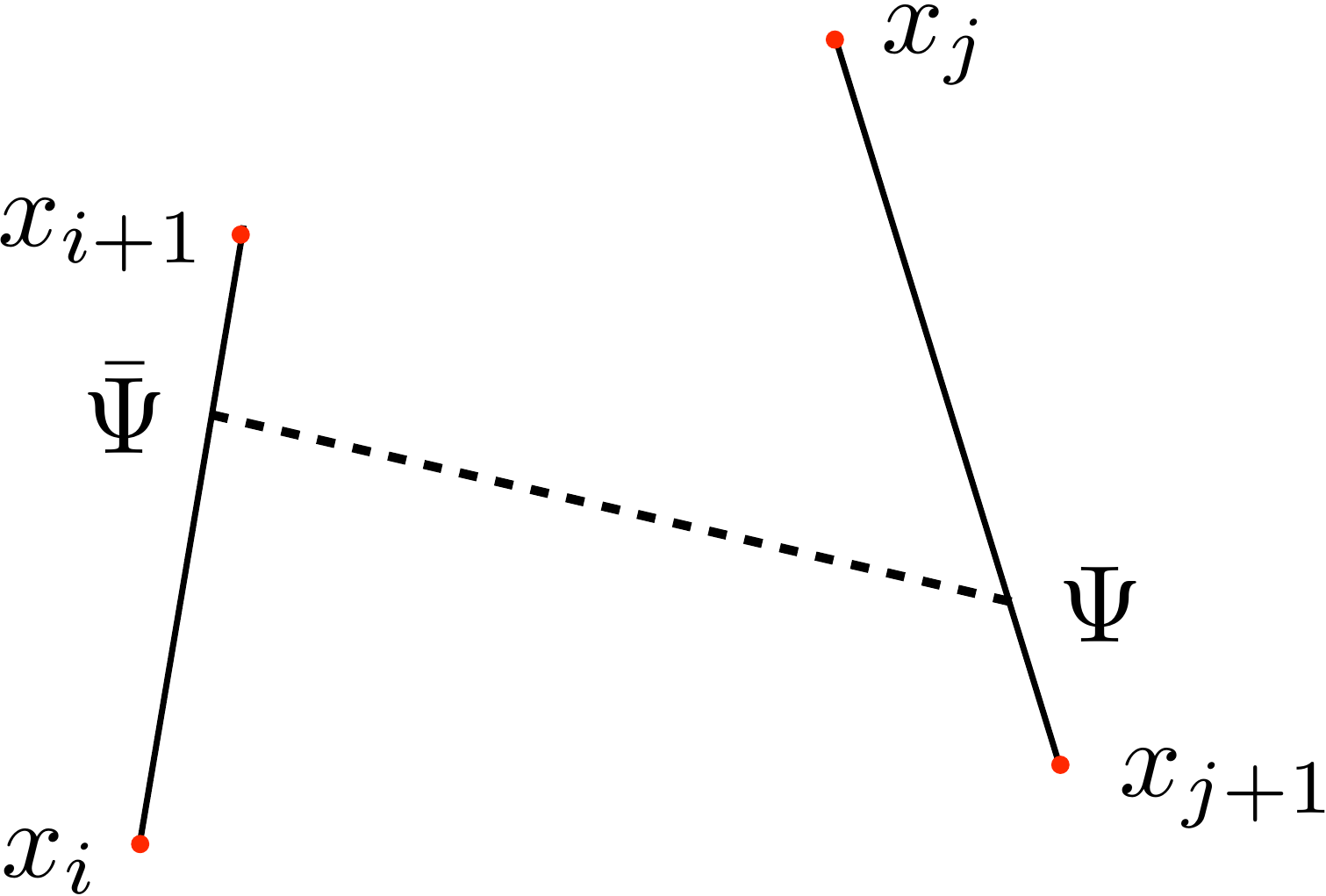}
	\caption{\it\small In the Abelian theory at order ${\rm g}^2$, the $\bar Q^{(1)}$ variation receives contributions only from a 
	single fermion propagator stretched between the two copies of the framed loop. In~\cite{Caron-Huot:2011ky}, Caron-Huot 
	showed that exactly this diagram resides at order $\chi\bar\chi$ in the non-chiral extension of the supersymmetric Wilson Loop. 
	Here we have discovered the same object purely within the chiral superloop dual to the superamplitude. }
	\label{fig:susycontract}
\end{figure}

Since we are only interested in the variation of the 1-loop MHV amplitude, it suffices to compute the correlator~\eqref{insertion} only to 
first non-trivial order in $\theta$, and only using the self-dual theory. To this order  the superconnection is simply
\be
	\mathbb{A}(x,\theta) 
	= A+ \im|\theta^a\ra[\bar\psi_a| + \cO(\theta^2)
\ee
and so the only possible contribution is from a single fermion exchanged between the two copies $C$ and $C'$ of the loop, as in figure~\ref{fig:susycontract}. Inserting the fermion propagator
\be
	\left\la \psi_\alpha^{\ a}(x)\bar\psi_{b\dot\beta}(y)\right\ra = \im\delta_b^{\ a} \frac{(x-y)_{\alpha\dot\beta}}{(x-y)^4}
\ee
and performing the integrals around both copies of the loop, the Ward identity gives to lowest order
\be
\begin{aligned}
	\sum_i\bar\epsilon\cdot \chi_i\frac{\del }{\del\mu_i}\rW^{\rm framed}[C_n]
	&=\sum_{i,j}\  
	\left\la\int_{x_i}^{x_{i+1}}\hspace{-0.2cm}[\bar\epsilon_a|\rd x|\psi^a(x)\ra 
	\int_{x_j}^{x_{j+1}}\hspace{-0.2cm}[\bar\psi_b(y)|\rd y|\theta^b\ra \right\ra
	 + \ (i\leftrightarrow j)\\
	&=\sum_{i,j}\  \int_0^1\!{\rm d}s \int_0^1\!{\rm d}t\ 
	\frac{[i\,\bar\epsilon_a]\,\chi_j^a \,\la i|x_{ij}|j]}{(x_{ij}+sx_{i+1\,i}-ty_{j+1\,j})^4} \ + \ (i\leftrightarrow j)\\
	&=\sum_{i,j} \frac{(i\!-\!1,i,i\!+\!1,\bar\epsilon_a)}{(i\!-\!1,i,i\!+\!1,j)}\,\chi_j^a \,
	\log \frac{x_{i\,j+1}^2 x_{i+1\,j}^2}{x_{ij}^2\, x_{i+1\,j+1}^2}\ + \ (i\leftrightarrow j)
\end{aligned}
\label{delta1WLresult}
\ee
where $(i,j,k,l)$ denotes the SL$(4;\C)$-invariant  contraction $\varepsilon_{ABCD}Z_i^AZ_j^BZ_k^CZ_l^D$. In this equation, $i$ and $j$ run around the two copies of the framed loop and the sum over $(i\leftrightarrow j)$ accounts for the fact that we could perform the $\bar Q^{(1)}$ variation on either copy. By construction $x_{ij}^2\neq0$ for all pairs of vertices $x_i$ and $x_j$ the summand in~\eqref{delta1WLresult} is always well-defined even in four dimensions.

The momentum twistor fermions $\chi_i$ may be varied independently, so we immediately deduce that at order $\theta^0$ the 1-loop symbol of this framed Wilson Loop is
\be
\begin{aligned}
	 \cS(\rW^{\rm framed}[C_n]|_{\rm g^2}) 
	 &= \sum_{i,j} \frac{X_i\!\cdot\! X_{j+1}\ X_{i+1}\!\cdot\! X_j}{X_i\!\cdot\! X_j\ X_{i+1}\!\cdot\! X_{j+1}}
	\ \otimes\ (i\!-\!1,i,i\!+\!1,j)\quad +\quad (i\leftrightarrow j)\\
	&=\sum_{i,j} X_i\!\cdot X_j \ \otimes\ \frac{(i\!-\!1,i,i\!+\!1,j\!-\!1)(i\!-\!2,i\!-\!1,i,j)}{(i\!-\!1,i,i\!+\!1,j)(i\!-\!2,i\!-\!1,i,j\!-\!1)}
	\quad + \quad (i\leftrightarrow j)\ .
\end{aligned}
\label{MHV1symbol}
\ee
This is the same as the symbol of the 1-loop cross-correlator of two (bosonic) Wilson Loops computed 
in~\cite{Gaiotto:2011dt}\footnote{Here, the twistor line $X_i$ is the line $(i,i\!-\!1)$ joining twistors $\cZ_i$ and $\cZ_{i-1}$ whereas 
in~\cite{Gaiotto:2011dt} $X_i$ denotes the line $(i\!+\!1,i)$.} from a gluon exchanged between the two loops. Since the 1-loop MHV 
amplitude may be computed by stretching a single gluon across the Wilson Loop, we have verified that the chiral superloop (or 
scattering amplitude) obeys the (dual) super Ward identity~\eqref{Wardid}, at least to 1-loop order in the MHV sector\footnote{A pleasing 
feature of the framing regularization is that all cases may be considered equally; there is a universal expression for the propagator 
between edges $i$ and $j$ which remains finite as the framing is removed if $|i-j|> 1$ and otherwise diverges logarithmically.}.

We emphasize that we do not expect this statement to strongly depend on the choice of regularization scheme. For example, in dimensional regularization, the four particle 1-loop MHV amplitude is given by~\cite{Green:1982sw}
\be
	\cM_\MHV^{n=4} = -\frac{2}{\epsilon^2}\left[\left(\frac{\mu^2}{-s}\right)^{\!\!\epsilon} +\left(\frac{\mu^2}{-t}\right)^{\!\!\epsilon}\right] 
	- \log^2\frac{s}{t}+\pi^2+\cO(\epsilon)\ ,
\label{4ptmassreg}
\ee
where $d=4-2\epsilon > 4$. Acting with $\bar Q_{\rm ext} = \sum \chi_i\del/\del\mu_i$ and ignoring order $\epsilon$ corrections gives 
\be
	[\bar\epsilon\cdot\bar Q_{\rm ext},\cM_\MHV^{n=4}] 
	= -\frac{1}{\epsilon}
	\left[\frac{(\bar\epsilon_a234)\chi_1^a+\hbox{cyclic}}{(1234)}
	\right]\left[\left(\frac{\mu^2}{-s}\right)^{\!\!\epsilon} +\left(\frac{\mu^2}{-t}\right)^{\!\!\epsilon} \right]\ .
\ee
One readily finds the same result by using the dimensionally regularized fermion propagator
\be
	\left\la\Psi(x)_\alpha^a\bar\Psi(y)_{\dot\alpha b}\right\ra = \delta^a_{\ b}\frac{(x-y)_{\alpha\dot\alpha}}{((x-y)^2+\im0)^{2-\epsilon}} \label{Coulombfermprop}
\ee
in~\eqref{delta1WLresult}. 

Finally, as an alternative method, we could have equally computed $\left[\bar\epsilon\!\cdot\!\bar Q_{\rm ext},\cM_{\rm MHV}\right]$ 
using only the self-dual supersymmetry transformations~\eqref{susy1}. The only part of the action that is not invariant under $\bar Q^{(0)}$ is $S_{\rm MHV}$, given in~\eqref{MHVaction}. In the Abelian case, this MHV action reduces to 
$\frac{\rm g^2}{2}\int \rd^4y\,G_{\beta\gamma}G^{\beta\gamma}$. Therefore, upon integrating by parts in the path 
integral we have
\be
\begin{aligned}
	\sum_i \chi_i\frac{\del}{\del\mu_i} \rW[C_n]
	&= -\im\left\la \left[\bar Q^{(0)}\,,S_{\rm MHV}\right]\,\exp\left(\im\oint_{C_n}\mathbb{A}\right)\right\ra\\
	&=-\im{\rm g^2}\int\rd^4y \left\la\del_{\dot\alpha\beta}\psi_\gamma^a(y)\, G^{\beta\gamma}(y)\,
	\exp\left(\im\oint_{C_n}\mathbb{A}\right)\right\ra\ .
\end{aligned}
\ee
A further integration by parts\footnote{The Wilson Loop lives in the conformally compactified space, so there is no boundary term.} 
transfers the $y$-derivative to $G$, producing a term proportional to the Abelian equation of motion for $G$. The correlation function  is then non-vanishing only when this equation of motion is localized to the Wilson Loop contour, and again leaves us with an insertion of $\psi$ on $C_n$.

%%%%%%%%%%%%%%%%%%%%%%%%%%%%%%%%%%%%%%%%%%%%%%%%%%%
%%%%%%%%%%%%%%%%%%%%%%%%%%%%%%%%%%%%%%%%%%%%%%%%%%

\section{Descent equations}
\label{sec:descent}

In the Abelian case, it was straightforward to compute the right hand side $\la[\bar Q^{(1)},{\rm W}]\ra$ of the Ward identity directly. 
However, calculating this correlation function was really no simpler than the original computations 
of~\cite{Drummond:2007aua,Brandhuber:2007yx} for the 1-loop MHV amplitude itself. To understand the anomaly in 
$\bar Q_{\rm ext}$ beyond one loop, or beyond the MHV sector, we must consider the non-Abelian theory. A direct computation then 
becomes even less appealing, both because the variation of the non-Abelian superconnection is more complicated, and because many 
more diagrams contribute.

To do better, in this section we will reformulate the right hand side of the Ward identity as an operation that may be carried out purely at 
the level of the external twistor data. The resulting equation may be interpreted as a descent equation that controls the 
structure of the dual supersymmetry $\bar Q$ anomaly.

\medskip

In the non-Abelian case, the superconnection $\mathbb{A}$ is determined by the constraints
\be
\begin{aligned}
	\lambda^\alpha\lambda^\beta \left[ D_{\alpha\dot\alpha}\,,D_{\beta b}\right] &= 0\\
	\lambda^\alpha\lambda^\beta\left\{D_{\alpha a}\,,D_{\beta b}\right\} &=0
\end{aligned}
\label{integrability}
\ee
that state that $\mathbb{A}$ is integrable along super null rays $\cong \R^{1|4}$. Up to order $\theta^4$, these constraints are solved by~\cite{Mason:2010yk,Belitsky:2011zm}
\be
\begin{aligned}
	A(x,\theta) &= A(x) + \im|\theta^a\ra[\bar\psi_a| + \frac{\im}{2}|\theta^a\ra\la\theta^b|D\phi_{ab} 
	-\frac{1}{3!}\varepsilon_{abcd}|\theta^a\ra\la\theta^b|\,D\la\theta^c\psi^d\ra\\
	&\hspace{0.5cm} +\frac{1}{4!}\varepsilon_{abcd}|\theta^a\ra\la\theta^b|\,D\la\theta^c|G|\theta^d\ra + \cdots\\
	|\Gamma_a(x,\theta)\ra &= \frac{\im}{2}\phi_{ab}|\theta^b\ra -\frac{1}{3}\varepsilon_{abcd}|\theta^b\ra\la\theta^c\psi^d\ra 
	+\frac{\im}{8}\varepsilon_{abcd}|\theta^b\ra\la\theta^c|G|\theta^d\ra+\cdots\ .
\end{aligned}
\label{superconncompts}
\ee
Note that the fermionic component of the superconnection $|\Gamma_a\ra$ depends only on the component fields $\{\phi,\psi,G\}$ and 
so is unaffected by the non-Abelian $\bar Q^{(1)}$ transformation
\be
	\delta^{(1)} \!A = \im |\psi^a\ra[\bar\epsilon_a| \qquad\qquad
	\delta^{(1)} |\bar\psi_a] = -\frac{\im}{2}\,|\bar\epsilon_c]\left[\phi^{cb},\phi_{ba}\right]\ .
\ee
Therefore, in the non-Abelian case we find that
\be
	\delta^{(1)}\!\left[{\rm Tr\,P}\exp\left(\im\oint_{C_n} {\mathbb A}\right)\right]
	=\im\,{\rm Tr} \left(\oint\delta^{(1)}\!\mathbb{A}(x,\theta)\ {\rm Hol}_{(x,\theta)}[\mathbb{A};C_n]\right)
\label{NonAbspacetimevary}
\ee
where
\be
\begin{aligned}
	\delta^{(1)}\!\mathbb{A} &= \left(\im|\psi^a\ra[\bar\epsilon_a|+\frac{1}{2}|\theta^a\ra[\bar\epsilon_c|\left[\phi^{cb},\phi_{ba}\right]
	+\frac{\im}{2}|\theta^a\ra [\bar\epsilon_e|\left[\la\theta^b\psi^e\ra,\phi_{ab}\right]+\cdots\right)\rd x\\
	&\hspace{0.5cm}+\varepsilon_{abcd}\left(-\frac{1}{3!}|\theta^a\ra[\bar\epsilon_e|\left[\la\theta^b\psi^e\ra,\la\theta^c\psi^d\ra\right]
	+\frac{\im}{4!}|\theta^a\ra[\bar\epsilon_e|\left[\la\theta^b\psi^e\ra,\la\theta^c|G|\theta^d\ra\right]+ \cdots\right)\rd x\ .
\end{aligned}
\label{nonAbvarycompts}
\ee
As at the end of the previous section, an alternative way to arrive at the same result is again to note that
\be
	\sum_i\chi_i\frac{\del}{\del\mu_i}\rW = \left\la\left[\bar\epsilon\!\cdot\!\bar Q^{(0)},\rW[C_n]\right]\right\ra
	=-\im \left\la\left[\bar\epsilon\!\cdot\!\bar Q^{(0)},S_{\rm MHV}\right]\,\rW[C_n]\right\ra\ .
\ee
In the non-Abelian case, the $\bar Q^{(0)}$ variation of $S_{\rm MHV}$ is, schematically,
\be
	\delta^{(0)}S_{\rm MHV} \sim \int \rd^4y\,{\rm Tr}\left(\psi^a \bar\epsilon_a\times[\hbox{eom for $G$}] 
	+ \bar\epsilon_c\left[\phi^{cb},\phi_{ba}\right]\times [\hbox{eom for $\psi^a$}]\right)\ .
\label{NonAbeoms}
\ee
This insertion of the $G$ and $\psi^a$ equations of motion would vanish in the absence of any further operator insertions, but fails to 
vanish because of the Wilson Loop, where it becomes localized. Since $G$ is conjugate to $A$ and $\psi^a$ is 
conjugate to $\bar\psi_a$, the net effect is to insert a copy of the superconnection $\mathbb{A}$ on the Wilson Loop contour, where 
every occurrence of $A$ and $\bar\psi$ is replaced by $\psi$ and $[\phi,\phi]$, respectively. This is equivalent to an insertion of the 
$\bar Q^{(1)}$ transformation~\eqref{nonAbvarycompts}.

\medskip

\begin{figure}[t]
	\centering
	\includegraphics[height=45mm]{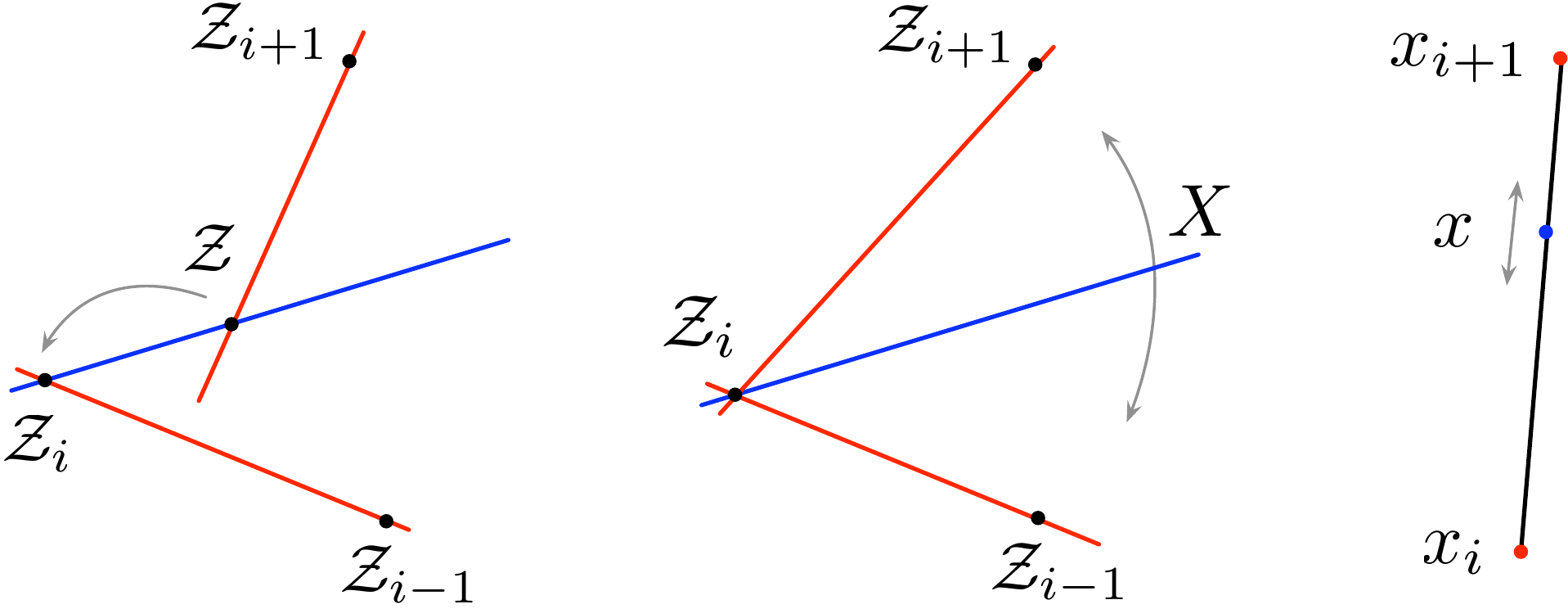}
	\caption{\it\small The $(n\!+\!1)$-point superloop is integrated over a contour that fixes $\cZ\to\cZ_i$ and also causes the line X
	to move in the plane $(i\!-\!1,i,i\!+\!1)$ between the lines $(i\!-\!1,i)$ and $(i,i\!+\!1)$. This corresponds to a point $x$ that is 
	integrated along the edge of the space-time Wilson Loop between $x_i$ and $x_{i+1}$.}
	\label{fig:contour}
\end{figure}

Inserting the expansion~\eqref{nonAbvarycompts} into~\eqref{NonAbspacetimevary} and directly computing the resulting 
correlator is clearly not the way to proceed. However, such Wilson loop correlators with operator insertions integrated along edges often arise from deformations of the contour, for example, collinear limits play an important role in the Wilson loop OPE. We wish to show that, instead of involving an unknown correlator, the super Ward identity~\eqref{Wardid} can be reformulated as
\be
	\sum_i\bar\epsilon\!\cdot\!\chi_i\frac{\del}{\del\mu_i}\rW[C_n]
	={\rm g}^2\sum_i \int\limits_{[0,\infty]\times S^1}\hspace{-0.2cm}
		V\!\lrcorner\rD^{3|4}\cZ\ \rW_{n+1}(\ldots,i,\cZ,i\!+\!1,\ldots)\ ,
\label{extWardid}
\ee
in terms of an $(n\!+\!1)$-point superloop. Here, $V=\bar\epsilon_a^{\ \dot\alpha}\chi^a\del/\del\mu^{\dot\alpha}$ is the vector field on 
twistor space that generates the usual $\bar Q^{(0)}$ transformations, while $\rD^{3|4}\cZ$ is the standard holomorphic measure on 
the Calabi-Yau superspace~\cite{Witten:2003nn} so that
\be
	V\!\lrcorner\rD^{3|4}\cZ = (\bar\epsilon_a,Z,\rd Z,\rd Z) \,\rd^4\chi \ \chi^a\ .
\label{Vmeasure}
\ee
The Grassmann integral should be performed with the new $\chi$ treated as independent of the other $\chi_i$s. Bosonically, the 
additional twistor is constrained to lie in the plane $(i\!-\!1,i,i\!+\!1)$ and so may be parametrized as
\be
	Z = Z_i + p(Z_{i-1}+qZ_{i+1})
\label{paramZ}
\ee
whereupon the bosonic part of the measure becomes $(\bar\epsilon_a,i\!-\!1,i,i\!+\!1)\,\rd q\, p\rd p$. With this parametrization, the contour 
extracts the residue of the integrand at $p=0$, and integrates $q$ from $0$ to $\infty$ {}\footnote{The coordinate $q$ is related to the 
space-time parametrization $x(t) = x_i + t(x_{i+1}-x_i)$ by $$t = \frac{q\la i\!+\!1\,i\ra}{\la i\!-\!1\,i\ra-q\la i\!+\!1\,i\ra}\ .$$ In the Lorentzian 
case, $Z$ should lie in the intersection of $(i\!-\!1,i,i\!+\!1)$ with ${\PN:=\{Z\in\CP^{3}\, |\, Z\cdot\bar Z=0\}}$, where the dot 
implies the SU$(2,2)$ metric appropriate for the Lorentzian conformal group. In this case the $q$ contour should be $0\leq|q|\leq\infty$ in the direction $\arg(q)=\frac{\im}{2}\log\left( - Z_{i-1}\!\cdot\!\bar Z_{i+1}/Z_{i+1}\!\cdot\!\bar Z_{i-1}\right)$.}. This contour ensures that the line $(i\cZ)$ corresponds to the insertion point of $\delta^{(1)}\!\mathbb{A}$ in space-time; see figure~\ref{fig:contour}.

The importance of equation~\eqref{extWardid} is that it provides us with a representation of the action of $\bar Q^{(1)}$ on the external 
data. We have written~\eqref{extWardid} in the coupling constant normalization~\eqref{rescalefields} adapted to agree with perturbation 
theory of the amplitude, making it clear that the formula is recursive; W$_{n+1}$ only needs to be known to order g$^{2\ell-2}$ in order to 
know the left hand side to order g$^{2\ell}$. More precisely, the differential of the Wilson Loop on the left hand side lowers its 
transcendentality by one, while on the right the integral $\rd q$ over the contour with boundary increases the transcendentality by one 
(both the Cauchy pole and the Grassmann integration preserve transcendentality, as they may be performed on the rational Yangian invariants - leading singularities - in front of the loop integrals). In this way the fact that $\ell$-loop amplitudes have transcendentality only $2\ell$, in accordance with the Kotikov-Lipatov principle~\cite{Kotikov:2002ab}, is made manifest.

On the other hand, the Grassmann degree of the left hand side is increased by one, while that of the right is decreased by three. Consequently, contributions to the $\bar Q^{(0)}$ variation of an N$^k$MHV amplitude are compensated by the $\bar Q^{(1)}$ transformation of an N$^{k+1}$MHV amplitude.

Putting these observations together and recalling that
\be
	(\bar Q^{(0)})^2 = (\bar Q^{(1)})^2 = 0 
	\qquad\qquad\hbox{and}\qquad\qquad 
	\bar Q^{(0)}\bar Q^{(1)}+\bar Q^{(1)}\bar Q^{(0)}=0\ ,
\ee
we see that we can view the Ward identity~\eqref{Wardid} or~\eqref{extWardid} as a descent equation that governs the anomaly in 
$\sum_i\chi_i\del/\del\mu_i$. Following the usual argument, since $\bar Q^{(0)}\cM^{\rm tree}= 0$ we have 
\be
	\bar Q^{(0)} \bar Q^{(1)}\cM^{\rm tree} = -\bar Q^{(1)}\bar Q^{(0)}\cM^{\rm tree} =0\ ,
\ee
so that\footnote{We assume that $\bar Q^{(0)}$ has trivial cohomology, at least at MHV level $k+\frac{1}{4}$ with $k\in\mathbb{Z}$.} 
$\bar Q^{(1)}\cM^{\rm tree}= \bar Q^{(0)} \widetilde\cM$ for some $\widetilde\cM$, which in~\eqref{extWardid} is identified as the 
1-loop amplitude $\widetilde\cM=\cM^{\rm 1-loop}$. Continuing the descent procedure generates (derivatives of) higher loop amplitudes. 

\medskip

Before demonstrating that~\eqref{extWardid} is equivalent to~\eqref{NonAbspacetimevary}, let us first gain some familiarity with it by 
recovering the result of section~\ref{sec:1loopMHV}. Up to order g$^2$ and $\theta$, the descent equation gives
\be
	\sum_i\bar\epsilon\!\cdot\!\chi_i\frac{\del}{\del\mu_i} \rW_{\rm MHV}^{\rm 1-loop}(1,\ldots,n)
	= \sum_i\int V\!\lrcorner\rD^{3|4}\cZ\  \rW_{\rm NMHV}^{\rm tree}(\ldots,i,\cZ,i\!+\!1,\ldots)\ .
\label{firstex}
\ee
Using the momentum twistor MHV expression
\be
	\rW_{\rm NMHV}^{\rm tree} = \sum_{k<j}\ [*,k,k\!+\!1,j,j\!+\!1]
\label{NMHVtreeCSW}
\ee
of the NMHV tree amplitude~\cite{Bullimore:2010pj} (corresponding to a single twistor superpropagator stretched across the twistor 
superloop~\cite{Mason:2010yk}), we note that the only R-invariants in this sum that have a pole as $\cZ\to\cZ_i$ are $[*,i,\cZ,j,j\!+\!1]$ for 
some $j$. We thus find
\be
\begin{aligned}
	&\sum_i\int\!V\!\lrcorner\rD^{3|4}\cZ \ \rW_{\rm NMHV}^{\rm tree}(\ldots,i,\cZ,i\!+\!1,\ldots)
	= \sum_{i,j}\int\!V\!\lrcorner\rD^{3|4}\cZ\ [*,i,\cZ,j,j\!+\!1]\\
	&=\sum_{i,j}\int(\bar\epsilon_a,Z,\rd Z,\rd Z)(j,j\!+\!1,*,i)^2\ 
	\frac{\chi_i^a(Z,j,j\!+\!1,*)+\chi_j^a(j\!+\!1,*,i,Z)+\chi_{j+1}^a(*,i,Z,j)}{(i,Z,j,j\!+\!1)\,(Z,j,j\!+\!1,*)\,(j\!+\!1,*,i,Z)\,(*,i,Z,j)}\\
	&=\sum_{i,j}\int\rd q\ (\bar\epsilon_a,i\!-\!1,i,i\!+\!1)\!\left[\frac{\chi_j^a(i,j,j\!+\!1,*)(j,j\!+\!1,*,i)}{(Z(s),j,j\!+\!1,*)\,(*,i,Z(s),j)}
	+\frac{ \chi_{j+1}^a(i,j,j\!+\!1,*)(j,j\!+\!1,*,i)}{(Z(s),j,j\!+\!1,*)\,(j\!+\!1,*,i,Z(s))}  \right]
\end{aligned}
\label{sumR}
\ee
where in going to the second line we have performed the Grassmann integral\footnote{We assume for simplicity that the reference 
supertwistor $\cZ_*=(Z_*,0)$; any non-vanishing $\chi_*$ may be verified to cancel around the sum.}, and in going to the third we 
used the explicit parametrization~\eqref{paramZ} and performed the $p$ contour integral. (In doing this, note that the $\chi_i$ 
term has a double pole at $p=0$, but no residue.) Collecting terms proportional to $\chi_j$ gives
\be
	\sum_{i,j}\int\rd q\ \frac{(\bar\epsilon_a,i\!-\!1,i,i\!+\!1)\,(i,j\!-\!1,j,j\!+\!1)}{(i,Z(s),j,j\!+\!1)\,(i,Z(s),j\!-\!1,j)}\chi_j^a
	=\sum_{i,j} \frac{(\bar\epsilon_a,i\!-\!1,i,i\!+\!1)}{(j,i\!-\!1,i,i\!+\!1)} \,
	\log\frac{X_i\!\cdot\! Y_{j+1}\ X_{i+1}\!\cdot\!Y_j}{X_i\!\cdot\!Y_j\ X_{i+1}\!\cdot\! Y_{j+1}}
\ee
in agreement with~\eqref{delta1WLresult}\footnote{Here we have been a little cavalier with the regularization. One may check that the 
exact expression~\eqref{delta1WLresult}, including the  $i\leftrightarrow j$ term, is reproduced from the descent equations for the framed 
superloop~\eqref{framed}.}.

\medskip

We now turn to relating our descent equation to the space-time correlator~\eqref{NonAbspacetimevary}. To do so, we must recall a few 
facts about the twistor space formulation of the superloop. In~\cite{Mason:2010yk,Bullimore:2011ni} it was shown that the superloop 
could be defined in twistor space as a product of holomorphic frames around the nodal curve 
$(\cZ_1\cZ_2)\cup(\cZ_2\cZ_3)\cup\cdots\cup(\cZ_n\cZ_1)$ that corresponds to the null polygon $C_n$ in space-time. (We will also call 
this twistor curve $C_n$.) The holomorphic frame $h(x,\theta;\lambda)$ is a smooth gauge transformation that defines a holomorphic 
trivialization of the twistor gauge bundle over the line X; {\it i.e.}, $h^{-1}\circ(\delbar-\im\cA)|_{\rm X}\circ h = \delbar|_{\rm X}$, so that $h$ 
obeys 
\be
	(\delbar-\im\cA)|_{\rm X} h=0
\label{hdef}
\ee
where $\delbar|_{\rm X}$ is the $\delbar$-operator with respect to 
$\lambda$. 

For a given $\cA$, equation~\eqref{hdef} uniquely defines the frame only up to gauge transformations $h(x,\theta;\lambda)\to   h(x,\theta;\lambda)g(x,\theta) $, where $g$ must be globally holomorphic in, and hence independent of, $\lambda$. In particular, we can use this freedom to pick a frame
\be
	\rU_{\rm X}(\cZ,\cZ_i) := h(x,\theta;\lambda) h^{-1}(x,\theta;\lambda_i)
\label{Udef}
\ee
that is normalised to be the identity at a some point $\cZ_i$ on the line X. The twistor space Wilson Loop used in~\cite{Mason:2010yk,Bullimore:2011ni} is then the product
\be
	\rW[C_n] = \frac{1}{N}\left\la {\rm Tr}\left(\cdots \rU_{{\rm X}_{i+1}}(\cZ_{i+1},\cZ_i)\rU_{\rm X_{i+1}}(\cZ_i,\cZ_{i-1})\cdots\right)\right\ra
\ee
of these holomorphic frames around $C_n$. This product computes a complex analogue of the trace of the holonomy of the  
partial, or (0,1), connection $\delbar-\im\cA$ around the holomorphic curve in twistor space.

The space-time superconnection may be recovered from these holomorphic frames in the standard way~\cite{Mason:2010yk,Sparling}. 
In particular, to recover $|\Gamma_a(x,\theta)\ra$ one first shows that, although $h$ itself depends smoothly on $\lambda$, the 
combination $\la\lambda\, \del_a h^{-1}\ra h$ is in fact globally holomorphic.  Since it clearly has homogeneity $+1$, Liouville's theorem 
implies that it must be linear, so that
\be
	\lambda^\alpha\frac{\del h^{-1}}{\del\theta^{\alpha a}} \,h  = \im \lambda^\alpha\Gamma_{\alpha a}(x,\theta)\ ,
\label{Gammafromh}
\ee
for some field $\Gamma_{\alpha a}$ that depends only on space-time. If we use this field to define a fermionic covariant derivative 
$\la\lambda D_a\ra = \la\lambda| \del_a-\im\Gamma_a\ra$ projected along $|\lambda\ra$, then~\eqref{Gammafromh} 
immediately implies that 
$D_{\alpha a}$ satisfies the integrability constraint in~\eqref{integrability}, from which the full $\Gamma_{\alpha a}$ may be 
reconstructed. In addition, multiplying both~\eqref{Gammafromh} and its bosonic counterpart on the right by $h^{-1}$, we see that 
$h^{-1}$ also obeys the defining equation for the super null Wilson Line in space-time. They must thus agree, up to a gauge transformation.

In fact, by pairing the holomorphic frames differently around the curve as
\be
	 \cdots 
	\lefteqn{
	\overbrace{\phantom{h(x_{i+1};\lambda_{i+1})\,h^{-1}(x_{i+1},\lambda_i)}}^{\textstyle{\rU_{{\rm X}_{i+1}}(\cZ_{i+1},\cZ_i)}}
	}
	h(x_{i+1};\lambda_{i+1})\,
	\underbrace{h^{-1}(x_{i+1},\lambda_i)\,h(x_i,\lambda_i)}_{\textstyle{{\rm P}\exp\im\int_{x_i}^{x_{i+1}}\!\!\mathbb{A}}}
	\lefteqn{
	\hspace{-1.5cm}\overbrace{\phantom{h(x_i,\lambda_i)\,h^{-1}(x_i,\lambda_{i-1})}}^{\textstyle{\rU_{{\rm X}_i}(\cZ_i,\cZ_{i-1})}}
	}
	\underbrace{h^{-1}(x_i,\lambda_{i-1})\,h(x_{i-1},\lambda_{i-1})}_{\textstyle{{\rm P}\exp\im\int_{x_{i-1}}^{x_i}\!\!\mathbb{A}}}
	\cdots\ ,
\label{bothloops}
\ee
both the twistor and space-time superloops may be exhibited simultaneously. If we use this observation in~\eqref{NonAbspacetimevary} 
we find that
\be
\begin{aligned}
	\delta^{(1)}\rW
	&= \frac{\im}{N}\oint_{C_n}\rd x^{\alpha\dot\alpha} \ {\rm Tr\,P}\left[ 
	\cdots \exp\left(\im\int_x^{x_{i+1}}\!\!\!\!\mathbb{A}\right)\,\delta^{(1)}\!\mathbb{A}_{\alpha\dot\alpha}(x)\,
	\exp\left(\im\int_{x_i}^x\!\!\mathbb{A}\right)\cdots\right]\\
	&= \frac{\im}{N} \oint\rd x^{\alpha\dot\alpha} \ {\rm Tr}\left[\cdots h^{-1}(x_{i+1},\lambda_i)\,h(x,\lambda_i)\,
	\delta^{(1)}\!\mathbb{A}_{\alpha\dot\alpha}(x)\,h^{-1}(x,\lambda_i)\,h(x_i,\lambda_i)\cdots
	\right]\\
	&=\frac{\im}{N}\oint\rd x^{\alpha\dot\alpha} \ {\rm Tr}\left[\cdots\rU_{{\rm X}_{i+1}}(\cZ_{i+1},\cZ_i)\,
	h(x,\lambda_i)\,\delta^{(1)}\!\mathbb{A}_{\alpha\dot\alpha}(x)\,h^{-1}(x,\lambda_i)\,
	\rU_{{\rm X}_i}(\cZ_i,\cZ_{i-1})\cdots\right]
\end{aligned}
\label{varywithh}
\ee
so that all the holomorphic frames except those immediately adjacent to the insertion may be paired into Us.

\medskip

Let us now relate this to the recursive formula~\eqref{extWardid}.  Writing the $(n\!+\!1)$-point superloop in terms of the expectation 
value of the product
\be
	\cdots \rU(\cZ_{i+1},\cZ)\rU(\cZ,\cZ_i)\rU(\cZ_i,\cZ_{i-1})\cdots
\ee
of holomorphic frames, we must carry out the integral over the additional supertwistor $\cZ$.  The only piece 
that depends on $\cZ$ is $\rU(i\!+\!1,\cZ)\rU(\cZ,i)$ and a little thought shows that the only contributions to the contour integral come from 
the Grassmann integrals acting on $\rU(\cZ,i)$, so that field insertions get trapped between $\cZ_i$ and $\cZ$ as we take the residue 
where $\cZ\to\cZ_i$.

Focusing on the Poincar{\'e} supersymmetry $\bar\epsilon^{\dot\alpha}_{\ a}\bar Q^a_{\ \dot\alpha}$, the bosonic measure in~\eqref{Vmeasure} becomes
\be
	(\bar\epsilon_a,Z,\rd Z,\rd Z) = [\bar\epsilon_a\rd \mu]\,\la\lambda\rd\lambda\ra  
	= [\bar\epsilon_a|\rd x|\lambda\ra\,\la\lambda\rd\lambda\ra\ .
\label{newmeasure}
\ee
To compute the effect of the Grassmann integration on $\rU(\cZ,i)$ we must replace 
the Grassmann integrals $\rd\chi \equiv \del/\del\chi$ by an operation on the $\theta$s, because $\rU_{\rm X}(\cZ,i)$ depends 
on these fermions only through its dependence on the line $(x,\theta)$.  If we pull back a function $f(\cZ)$ on super twistor 
space to the spin bundle by setting $|\mu]= x|\lambda\ra$ and $\chi=\theta|\lambda\ra$, then
\be
	\left.\left(\frac{\del f}{\del\chi^a}\right)\right|_{\chi=\theta\lambda} 
	= \frac{1}{\la\rho\lambda\ra} \rho^\alpha\frac{\del f|_{\chi=\theta\lambda}}{\del\theta^{\alpha a}}
\label{choiceoflift}
\ee
for an arbitrary reference spinor $|\rho\ra$ that defines a choice of lift of $\del/\del\chi$ to the spin bundle. By definition $\theta = (\chi_i\lambda - \chi\lambda_i)/\la\lambda i\ra$, so to ensure we do not pick up contributions from $\chi_i=\theta|i\ra$, we must choose the lift $|\rho\ra = |i\ra$, since $\la i\frac{\del}{\del\theta}\ra\,\chi_i = \la i\frac{\del}{\del\theta}\ra\, \theta|i\ra =0$.

From the defining equation~\eqref{hdef} we see that
\be
\begin{aligned}
	0&=\frac{\del}{\del\theta^{\alpha a}}\left[\phantom{\frac{1}{1}}\hspace{-0.2cm}
	(\delbar -\im \cA)|_{\rm X} \rU_{\rm X}(\cZ,\cZ_i)\right] \\
	&= (\delbar -\im \cA)\,|\del_a\rU_{\rm X}(\cZ,\cZ_i) \ra
	-\im|\lambda\ra\frac{\del\cA}{\del\chi^a}\, \rU_{\rm X}(\cZ,\cZ_i)\ ,
\end{aligned}
\ee
where we understand that the $(\delbar-\im\cA)$-operator is always pulled back to X, so that in particular $\cA$ depends on $\theta$ only through $\chi=\theta|\lambda\ra$. Given that $\rU$ solves~\eqref{hdef}, this is solved in terms of the integral
\be
	|\del_a\rU_{\rm X}(\cZ,\cZ_i) \ra
	=\im\int_{\rm X}\frac{\la\lambda'\rd\lambda'\ra\la\lambda\,i\ra}{\la\lambda\lambda'\ra\la\lambda'i\ra}
	\rU_{\rm X}(\cZ,\cZ')\,|\lambda'\ra\frac{\del\cA}{\del{\chi'}^a}\,\rU_{\rm X}(\cZ',\cZ_i)
\label{Uderiv}
\ee
over another point $\cZ'\in{\rm X}$. 

The fermionic measure $\rd^4\chi\,\chi^a$ means that we never get contributions from the term of order $\chi^4$ in the holomorphic frame $\rU_{\rm X}(\cZ,\cZ_i)$. However, it is convenient to temporarily include such contributions and then later project them out. Thus, combining equations~\eqref{newmeasure},~\eqref{choiceoflift} and~\eqref{Uderiv} we consider the following expression involving three fermionic derivatives
\be
\begin{aligned}
	&\int(\bar\epsilon_a,Z,\rd Z,\rd Z)\frac{\varepsilon^{abcd}}{3!}\frac{\del^3}{\del\chi^b\del\chi^c\del\chi^d} \rW(\cdots,i,\cZ,i\!+\!1,\cdots)
	\\
	&=-\frac{\im}{3! N}\int [\bar\epsilon_a|\rd x|\lambda\ra \la\lambda\rd\lambda\ra\, 
	{\rm Tr}\left[\cdots \varepsilon^{abcd}\frac{\la i\del_b\ra\la i\del_c\ra }{\la i\lambda\ra^2}
	\int_{\rm X}\frac{\la\lambda'\rd\lambda'\ra}{\la\lambda\lambda'\ra}
	\rU_{\rm X}(\cZ,\cZ')\,\frac{\del\cA}{\del{\chi'}^d}\,\rU_{\rm X}(\cZ',\cZ_i)\cdots\right]\\
	&=-\frac{\im}{3! N}\int_{x_i}^{x_{i+1}}\hspace{-0.2cm} \rd t\oint \la\lambda\rd\lambda\ra\, [\bar\epsilon_a \tilde \imath] \ 
	{\rm Tr}\left[\cdots \varepsilon^{abcd}\frac{\la i\del_b\ra\la i\del_c\ra }{\la i\lambda\ra}
	\int_{\rm X}\frac{\la\lambda'\rd\lambda'\ra}{\la\lambda\lambda'\ra}
	\rU_{\rm X}(\cZ,\cZ')\,\frac{\del\cA}{\del{\chi'}^d}\,\rU_{\rm X}(\cZ',\cZ_i)\cdots\right]\ ,
\label{3rdderiv}
\end{aligned}
\ee
where in going to the last line we used the fact that $\rd x = |\tilde \imath]\la i|\, \rd t$ on the $i^{\rm th}$ edge. It is now straightforward to perform the contour integral setting $\la i\lambda\ra=0$, which leaves us with
\be
	-\frac{\im}{N} \int_{x_i}^{x_{i+1}}\hspace{-0.2cm} \rd t\,[\bar\epsilon_a \tilde \imath] \ 
	{\rm Tr}\left[\cdots \frac{\varepsilon^{abcd}}{3!}\la i\del_b\ra\la i\del_c\ra 
	\int_{\rm X}\frac{\la\lambda'\rd\lambda'\ra}{ \la i \lambda'\ra}
	\rU_{\rm X}(\cZ_i,\cZ')\,\frac{\del\cA}{\del{\chi'}^d}\,\rU_{\rm X}(\cZ',\cZ_i)\cdots\right]
\label{residue}
\ee
as a residue.

We can make sense of this expression if we note that, since $\rU(\cZ,\cZ_i)$ is normalized to be the identity when $\cZ=\cZ_i$, if we evaluate the equation
\be
	\la\lambda\Gamma_a(x,\theta) \ra= \la\lambda\,\del_a \rU^{-1}_{\rm X}(\cZ,\cZ_i)\ra\, \rU_{\rm X}(\cZ,\cZ_i) 
\label{GammaUigauge}
\ee
at $\cZ=\cZ_i$, the space-time connection $|\Gamma_a\ra$ defined in the gauge specified by these normalized holomorphic frames must obey $\la i\,\Gamma_a(x,\theta)\ra=0$. Hence {\it in this gauge} we have
\be
	|\Gamma_a(x,\theta)\ra = |i\ra \,\gamma_a(x,\theta;\lambda_i)
\label{littlegammadef}
\ee
where $\gamma_a$ is a fermionic Lorentz scalar that depends smoothly on $\lambda_i$ (the data of the gauge choice) as well as on $(x,\theta)$. Using~\eqref{Uderiv} in~\eqref{GammaUigauge} shows that
\be
	\gamma_a(x,\theta;\lambda_i) = \int_{\rm X}\frac{\la\lambda'\rd\lambda'\ra}{ \la i \lambda'\ra}
	\rU_{\rm X}(\cZ_i,\cZ')\,\frac{\del\cA}{\del{\chi'}^a}\,\rU_{\rm X}(\cZ',\cZ_i)\ ,
\label{littlegammaint}
\ee
exactly as appears in~\eqref{residue}.

The existence of a gauge in which $|\Gamma_a\ra = |\lambda\ra\gamma_a$ has a remarkable consequence that was also exploited in~\cite{Adamo:2011dq}. From the integrability conditions~\eqref{integrability}, the only non-vanishing part of the fermionic supercurvature is
\be
	\cW_{ab} = \im\epsilon^{\alpha\beta}\left\{D_{\alpha a}, D_{\beta b}\right\} 
	= \del^\alpha_{\ [a}\Gamma^{\phantom{a}}_{b]\alpha} -\im \{\Gamma^\alpha_{\ a},\Gamma_{\alpha b}\}
\ee
but if $|\Gamma_a\ra\propto|i\ra$ then the final anticommutator vanishes, even in the non-Abelian case. Furthermore, in this gauge we have
\be
\begin{aligned}
	\cW_{cd}&=\del^\alpha_{\  [c}\Gamma^{\phantom{a}}_{d]\alpha}=\la i\del_{\,[c}\ra\gamma_{d]}\\
	\lambda^\alpha_i D_{\alpha b}\cW_{cd} &= \la i\del_b\ra\cW_{cd}= \la i\del_b\ra\la i\del_{[c}\ra\gamma_{d]}\ .
\end{aligned}
\label{covderiv}
\ee
Using these together with~\eqref{littlegammaint} in~\eqref{residue}, we see that
\begin{multline}
	\frac{\varepsilon^{abcd}}{3!}\int(\bar\epsilon_a,Z,\rd Z,\rd Z)\frac{\del^3}{\del\chi^b\del\chi^c\del\chi^d} \rW(\cdots,i,\cZ,i\!+\!1,\cdots)
	\\
	=-\im\frac{\varepsilon^{abcd}}{3!N}\int_{x_i}^{x_{i+1}}\hspace{-0.2cm} \rd x^{\alpha\dot\alpha}
	\left\la{\rm Tr} \left[ \cdots \rU_{{\rm X}_{i+1}}(\cZ_{i+1},\cZ_i)\  \bar\epsilon_{a\dot\alpha}
	D_{\alpha b} \cW_{cd}\  \rU_{{\rm X}_i}(\cZ_i,\cZ_{i-1})\cdots \right]\right\ra
\label{5am}
\end{multline}
in terms of the covariant derivative of the fermionic supercurvature.

At present, the fields in~\eqref{5am} are expressed using a gauge that is natural on twistor space, but somewhat obscure on space-time. 
To obtain an expression purely in terms of space-time fields, we should transform to the gauge $(\delbar^\dagger\cA)_{\rm X}= 0$ where 
the twistor connection is harmonic (with respect to an arbitrary Hermitian metric) on the line X. The point of this gauge is that the 
$\lambda$-dependence of the twistor fields is completely fixed, so that they reduce to space-time fields 
(see~\cite{Boels:2006ir,Woodhouse:1985id} for details). We can assume that the basic holomorphic frame $h(x,\theta,\lambda)$ is 
chosen so as to put the fields in this gauge. Then since $\rU_{\rm X}(\cZ,\cZ_i) = h(x,\theta;\lambda)h^{-1}(x,\theta;\lambda_i)$, the 
definition~\eqref{GammaUigauge} of the space-time superconnection shows that if we replace 
$\rU_{\rm X}(\cZ,\cZ_i)\to\rU_{\rm X}(\cZ,\cZ_i)\,h^{-1}(x,\theta;\lambda_i) = h(x,\theta;\lambda)$, then the adjoint-valued derivative of the 
supercurvature in~\eqref{covderiv} transforms as
\be
	 D_{\alpha b}\cW_{cd} \to h(x,\theta,\lambda_i)\,D_{\alpha b}\cW_{cd}\,h^{-1}(x,\theta;\lambda_i)\ .
\ee
Comparing equations~\eqref{5am} \&~\eqref{varywithh}, we have shown that, once transformed to space-time, the expression in~\eqref{3rdderiv} is
\be
	\frac{\im}{N}\left\la{\rm Tr} \left(\oint \frac{\varepsilon^{abcd}}{3!} [\bar\epsilon_a|\rd x| D_b\cW_{cd}\ra\,{\rm Hol}_{(x,\theta)}[\mathbb{A};C_n]\right)\right\ra\ .
\ee
Before identifying this insertion with the variation $\delta^{(1)}\!\mathbb{A}(x,\theta)$ given in~\eqref{nonAbvarycompts}, there is one 
final subtlety to address. In replacing the Grassmann integral $\int\left(\rd^4\chi [\bar\epsilon_a| \chi^a\ \cdots\right)$ by the derivative 
$[\bar\epsilon_a| (\del^3/\del\chi^3)^a \left(\cdots\right)$, we should really set $\chi=0$ after taking the three derivatives. This merely 
expresses the fact that, due to the explicit $\chi$ in the measure, the superloop itself can only be expanded to order $(\chi)^3$. From the superconnection~\eqref{superconncompts} we find 
that
\be
\begin{aligned}
	\cW_{ab}(x,\theta) &= \im\phi_{ab}-\varepsilon_{abcd}\la\theta^c\psi^d\ra+\frac{\im}{2}\varepsilon_{abcd}\la\theta^c|G|\theta^d\ra 
	+ \frac{\im}{4}\left[\phi_{ac},\phi_{bd}\right]\la\theta^c\theta^d\ra + \cdots\\
	\frac{\varepsilon^{abcd}}{3!}|D_b\cW_{cd}\ra
	&= \im|\psi^a\ra+ G|\theta^a\ra
	+ \frac{1}{2}|\theta^a\ra\left[\phi^{cb},\phi_{ba}\right]+\cdots\ ,
\end{aligned}
\ee
which coincides with the $\bar Q^{(1)}$ variation $\delta^{(1)}\!\mathbb{A}$ in~\eqref{nonAbvarycompts} once we set to zero those 
components of the covariant derivative that originated from the $(\chi)^4$ term in the holomorphic frame; in the expansion above, this removes the term proportional to $G$. This may be achieved by inserting the projector
\be
{\mathbb{P}^e}_a = {\delta^e}_a - \frac{1}{4}\la \theta^e \del_a\ra\, .
\ee
We have thus demonstrated that 
\be
\begin{aligned}
	\sum_i\bar\epsilon\!\cdot\!\chi_i\frac{\del}{\del\mu_i}\rW[C_n]
	&=\im\,{\rm Tr} \left(\oint\delta^{(1)}\!\mathbb{A}(x,\theta)\ {\rm Hol}_{(x,\theta)}[\mathbb{A};C_n]\right)\\
	&=\im\,{\rm Tr} \left(\oint \frac{\varepsilon^{abcd}}{3!} [\bar\epsilon_e|\rd x| {\mathbb{P}^e}_a D_b\cW_{cd}\ra {\rm Hol}_{(x,\theta)}[\mathbb{A};C_n]\right)\\
	&={\rm g}^2\sum_i \int\limits_{[0,\infty]\times S^1}\hspace{-0.2cm} V\!\lrcorner\rD^{3|4}\cZ\ \rW(\ldots,i,\cZ,i\!+\!1,\ldots)\ ,
\end{aligned}
\label{descentfinal}
\ee
as claimed.

%%%%%%%%%%%%%%%%%%%%%%%%%%%%%%%%%%%%%%%%%%%%%%%%%%%
%%%%%%%%%%%%%%%%%%%%%%%%%%%%%%%%%%%%%%%%%%%%%%%%%%

\section{Interpretation for amplitudes}
\label{sec:exacting}

In this paper we have examined the failure of loop amplitudes to be annihilated by the dual superconformal generator 
$\bar Q= \sum_i\chi_i\del/\del\mu_i$. From the point of view of the dual superloop, this is an ordinary supercharge, and we have used 
this perspective throughout the paper.  However, it is also natural to wonder how our results, in particular the descent equation
\be
	\sum_i\chi_i\frac{\del}{\del\mu_i} \rW[C_n] = {\rm g}^2\sum_i\int V\!\lrcorner\rD^{3|4}Z\,\rW(\ldots,i,\cZ,i\!+\!1,\ldots)\ ,
\ee
arise from the point of view of scattering amplitudes, and how they are to be interpreted there.

They key to understanding this is to note that the particular dual Poincar{\'e} supercharge $\bar Q$ we have studied actually coincides with the original superconformal $\bar s$ supercharge~\cite{Drummond:2008vq}. This supercharge may be represented on the $n$-particle on-shell momentum space as 
\be
	\bar s= \sum_i\eta_i\frac{\del}{\del\bar\lambda_i}\ ,
\label{sbar}
\ee
where $\eta^a_i = (\chi^a_{i-1}\la i\,{i\!+\!1}\ra + \hbox{cyclic}) / \la i\!-\!1\, i\ra\la i\, i\!+\!1\ra$ in terms of the (momentum) twistor fermions 
$\chi_i$ we have used in the rest of the paper.  As shown 
in~\cite{Korchemsky:2009hm,Sever:2009aa,Bargheer:2009qu,Beisert:2010gn},  even the $n$-particle MHV tree amplitude is not strictly 
invariant under the transformations~\eqref{sbar} because of potential contributions to $\bar s$ at poles in $\lambda_i$ when external 
states become collinear. In our paper, we have assumed the initial superloop contour $C_n$ to be generic, so this tree-level failure is 
invisible.

\begin{figure}[t]
	\centering
	\includegraphics[height=45mm]{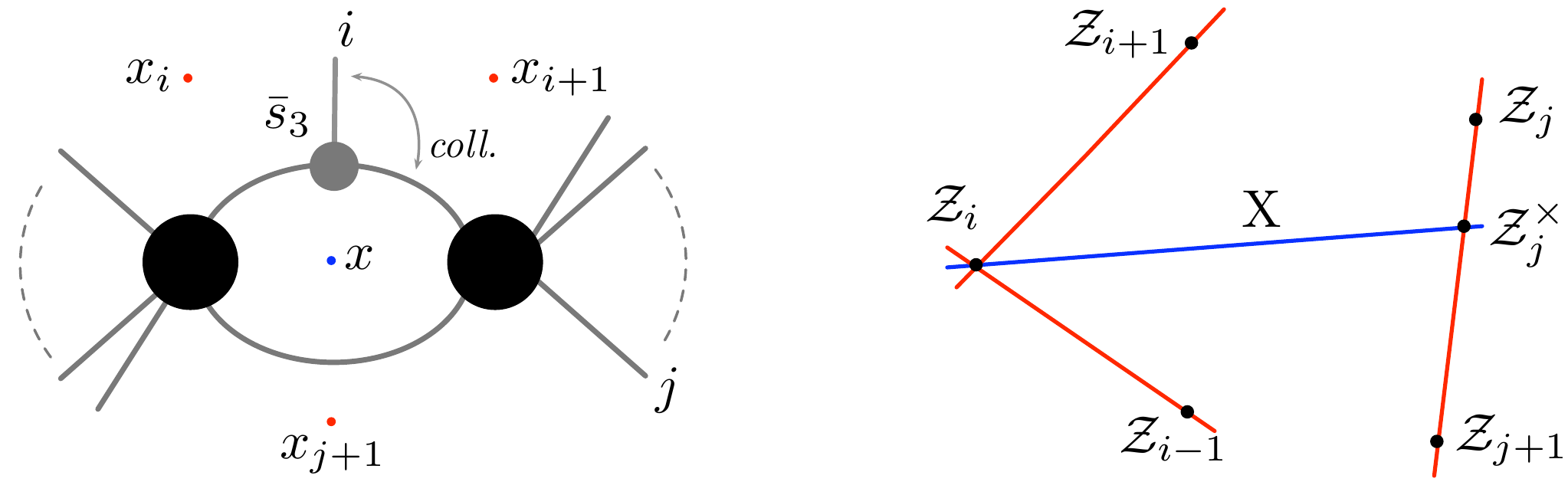}
	\caption{\it\small From the point of view of the amplitudes, corrections to the original superconformal generator $\bar s=\bar Q$ 
		arise when a loop momentum becomes collinear with some external momentum $p_i$. At one loop this contribution may be 
		represented by a dispersion integral of the cut diagram on the left. From the point of view of the superloop, the same 
		contribution arises as a particular BCFW decomposition of the $(n+1)$-point tree amplitude inside the descent equations.}
	\label{fig:collinearcut}
\end{figure}

However, \cite{Korchemsky:2009hm,Sever:2009aa,Beisert:2010gn} further showed that the same phenomenon is responsible for the 
violation of $\bar s$ by loop amplitudes, where the collinearity is now between an external momentum and a loop momentum. 
The authors of \cite{Sever:2009aa,Beisert:2010gn} further showed that the na\"ive action of~\eqref{sbar} on the loop amplitude could be 
`corrected' by deforming the way this generator acts on the fields. The required deformation 
includes\footnote{Beisert {\it et al.} also identified two other sources of contribution to the $\bar s$ anomaly in 
loops. The first arises because the amplitudes require regularization which inevitably breaks superconformal 
symmetry. However, it is less clear that one cannot find a regulator that preserves the Poincar{\'e} supersymmetry of the Wilson Loop (= 
dual supersymmetry of the amplitude). More practically, this contribution will vanish in any finite quantity such as the ratio function. The 
final potential contribution only arises if two or more external legs become collinear. We have ignored this possibility by our genericity 
assumption on $C_n$. It would clearly be interesting to revisit this issue.} a contribution that at 1-loop can be expressed as a dispersion 
integral of a rational function obtained from acting on the unitarity cut of the 1-loop amplitude in the limit that one of the cut loop 
propagators becomes collinear with one of the external momenta adjacent to it. See figure~\ref{fig:collinearcut} for an illustration. In this figure, the displayed propagators are understood to be cut, so that
\be 
		(x-x_i)^2=(x-x_{i+1})^2=(x-x_{j+1})^2=0\ .
\label{cuts}
\ee
The dispersion integral is over the fraction of momenta shared by the collinear states that is carried away by the external edge, while the 
operator $\bar s_3$ is the classical $\bar s$ operator acting on the three particle $\overline{\rm MHV}$ amplitude indicated in the figure 
(which is not zero in this collinear regime).

\medskip

To relate this to the descent equation~\eqref{descentfinal} for the superloop, we make use of a particular BCFW 
decomposition\footnote{In~\cite{Bullimore:2011ni} BCFW recursion for scattering amplitudes was identified with a particular version of 
the Migdal-Makeenko equations for Wilson Loops.} of the $(n\!+\!1)$-point superloop. Consider the BCFW deformation
\be
	\cZ \to \cZ(r)\equiv\cZ+ r\cZ_{i+1}
\ee
where $r$ is the deformation parameter. Applying this deformation to $\rW(\ldots,i,\cZ,i\!+\!1,\ldots)$ we see that as $r$ varies, the only 
twistor line to be affected is $(i\cZ)$. For the superloop in the self-dual theory (tree-level scattering amplitude) we have the BCFW 
decomposition
\begin{multline}
	\rW(\ldots,i,\cZ,i\!+\!1,\ldots) = \rW(\ldots,i,i\!+\!1,\ldots) \\
	+\  \sum_{j=i+2}^{i-2}\ [i,\cZ,i\!+\!1,j,j\!+\!1]\,\rW(j\!+\!1,\ldots,i,\cZ^\times_j)\,\rW(\cZ^\times_j,\cZ^\sharp_j,i\!+\!1,\ldots,j)\ ,
\label{BCFW}
\end{multline}
where $\cZ^\sharp_j$ is the value of $\cZ(r)$ when the deformed line intersects $(j\,j\!+\!1)$, and where $\cZ^\times_j$ is the intersection 
point. Using the fact that as $\cZ$ moves, it always lies in the plane $(i\!-\!1\,i\,i\!+\!1)$, this intersection point is simply the intersection\footnote{The condition that these lines and planes do intersect is non-trivial in the superspace. It is ensured by the R-invariant prefactor 
$[i,\cZ,i\!+\!1,j,j\!+\!1]$, which may be interpreted as a fermionic $\delta$-function with support only when its five arguments lie on a 
common $\CP^3\subset\CP^{3|4}$. See~\cite{Bullimore:2010pj,Adamo:2011pv} for further discussion.}
\be
	\cZ^\times_j = (i\!-\!1\,i\,i\!+\!1)\cap(j\,j\!+\!1)
\label{Zcross}
\ee
while
\be
	\cZ^\sharp_j = (\cZ\,i\!+\!1)\cap(j\,j\!+\!1\,i)
\label{Zsharp}
\ee
is the shifted point.

It is now easy to connect this to the amplitude picture discussed above. The first, `homgeneous' term in the recursion is independent of 
$\cZ$ and cannot contribute to the integral on the right hand side  of~\eqref{descentfinal}. For the other terms, both the Grassmann 
integrals and the contour integral setting $\cZ\to\cZ_i$ may be performed using the explicit R-invariant in~\eqref{BCFW}. Noting that the 
shifted point $\cZ^\sharp_j$ also reduces to $\cZ_i$ in this limit, we see that the integrand of the descent equation reduces to a product 
of two smaller Wilson Loops $\rW(j\!+\!1,\ldots,i,\cZ^\times_j)\rW(\cZ^\times_j,i,i\!+\!1,\ldots,j)$ that share the line $\rm X = (i\cZ_j^\times)$, 
times a coefficient that is left over from the R-invariant.

This situation is illustrated on the right of figure~\ref{fig:collinearcut}. The line X intersects the lines $(i\!-\!1,i)$, $(i,i\!+\!1)$ and $(j,j\!+\!1)$ 
and so obeys the cut conditions~\eqref{cuts}. This line corresponds to the location of the operator insertion 
$\delta^{(1)}\!\mathbb{A}(x,\theta)$ from the point of the superloop, and is the dual region momentum corresponding to the loop 
momentum in the amplitude. That X should be associated with a loop momentum in the scattering amplitude is particularly natural if we 
recall that quantum corrections to the amplitude correspond to insertions of $S_{\rm MHV}$ in the superloop, and that 
$\delta^{(1)}\mathbb{A}$ can be obtained from the $\bar Q^{(0)}$ transformation of $S_{\rm MHV}$ as discussed around~\eqref{NonAbeoms}. The kinematics of the three-particle $\overline{\rm MHV}$ amplitude -- all three particles sharing a common 
$\lambda$ spinor -- is also reflected in the figure as the triple intersection at $\cZ_i$.  The momentum in the cut propagator between 
$p_j$ and $p_{j+1}$ is determined in terms of $\cZ^\times_j$. Finally, the dispersion integral for the amplitude becomes the integral over 
the line X lying in the plane $(i\!-\!1,i,i\!+\!1)$, corresponding to integrating the insertion of $\mathbb{A}$ along the edge of the Wilson 
Loop.

We believe a similar story will be true for multi-loop amplitudes (or more precisely their finite ratio \& remainder functions) provided one 
uses the all-loop extension of the BCFW recursion relations discovered in~\cite{Arkani-Hamed:2010kv}.

%%%%%%%%%%%%%%%%%%%%%%%%%%%%%%%%%%%%%%%%%%%%%%%%%%%
%%%%%%%%%%%%%%%%%%%%%%%%%%%%%%%%%%%%%%%%%%%%%%%%%%

\vskip1.5truecm

\noindent{\Large\bf Acknowledgments}\\

It is a pleasure to thank Fernando Alday, Nathan Berkovits, Freddy Cachazo, Simon Caron-Huot, Amit Sever and 
especially Lionel Mason for helpful discussions.  MB is supported by an STFC Postgraduate Studentship. DS is supported by the 
Perimeter Institute for Theoretical Physics. Research at the Perimeter Institute is supported by the Government of Canada through 
Industry Canada and by the Province of Ontario through the Ministry of Research $\&$ Innovation.

\bibliographystyle{JHEP}
\bibliography{DualSusy}

\end{document}